\documentclass[conference]{IEEEtran}
\usepackage[T1]{fontenc} 
\usepackage{amsmath}
\usepackage[export]{adjustbox}
\usepackage{hyperref,comment}
\usepackage{graphicx}
\usepackage{subcaption}
\usepackage[dvipsnames]{xcolor}
\usepackage{algorithm}
\usepackage{algpseudocode}
\usepackage[utf8]{inputenc}
\usepackage{bm}
\usepackage{listings}
\usepackage{mathbbol}
\usepackage{dblfloatfix}

\author{Vasileios Georgiou \\ 
        Purdue University \\
        \texttt{vgeorgio@purdue.edu} \and 
        Petros Drineas \\
        Purdue University \\
        \texttt{pdrineas@purdue.edu}}

\newcommand{\ma}{\boldsymbol{A}}
\newcommand{\mb}{\boldsymbol{B}}
\newcommand{\mc}{\boldsymbol{C}}

\newcommand{\vb}{\boldsymbol{b}}

\newcommand{\vx}{\mathbf{x}}

\definecolor{dark-green}{RGB}{0,100,0}

\title{TOMOS: Eliminating Data Duplication in Mixed-Precision Matrix Computations}

\date{Spring $2025$}
\begin{document}
\maketitle
\noindent

\begin{abstract}
Mixed-precision algorithms exploit reduced-precision hardware, however computing with the same matrix at distinct precisions typically requires duplicate storage, repeated conversions, or operation-emulation. We introduce \textbf{TOMOS}\footnote{\url{https://github.com/vasilisge0/tomos}}, a composite-precision matrix format that provides implicit $\bm{32}$-bit and $\bm{16}$-bit access from a single representation. Each FP32 entry is split into a $\bm{16}$-bit primary word for low-precision computation and a complementary word that enables exact FP32 reconstruction. The encoding scheme improves numerical quality of the low-precision view through subnormal representations, and allows exact recovery of high-precision inputs.
TOMOS uses FP32-sized storage, with data movement scaling with compute precision. Optimized \texttt{GEMM} and \texttt{GEMV} kernels achieve throughput comparable to vendor libraries on NVIDIA A100 GPUs, while a mixed-precision least-squares solver achieves up to $\bm{1.6\times}$ speedup over FP32 without redundant matrix allocations.
\end{abstract}

\section{Introduction}

Modern accelerator-based systems deliver unprecedented peak arithmetic throughput, driven in large part by native support for reduced-precision floating-point formats. Specialized units operating on $16$-bit, $8$-bit, and even $4$-bit operands can provide orders-of-magnitude higher theoretical throughput than traditional double-precision computation. In large-scale machine learning and scientific computing workloads, however, arithmetic throughput alone does not determine performance: data movement, memory footprint, and memory management are often equally important. Consequently, realizing the benefits of reduced-precision hardware depends critically not only on the arithmetic format used for computation, but also on how numerical data is represented, stored, and accessed across multiple precisions.

In contemporary machine learning frameworks, mixed-precision execution has become standard practice \cite{Kalamkar2019BF16, micikevicius2018mixed, rakka2022mixed, schaefer2023mixed}. To preserve numerical stability, tensors frequently transition between precisions at predefined points in the computational graph. In practice, this often requires either storing multiple representations of the same data or repeatedly converting between formats. Both approaches increase memory footprint and memory traffic. Even when conversions are fused within kernels, they consume bandwidth, introduce additional instructions, and can create synchronization overheads. As a result, the cost of exposing multiple precision views of the same data can substantially reduce the performance gains offered by low-precision arithmetic.

Scientific computing presents a closely related challenge. Many numerical methods require high-precision guarantees to preserve stability and convergence, while selectively exploiting lower precision for performance. Techniques such as iterative refinement~\cite{Haidar2020, CarsonHigham2017, Dongarra2009} and error-free transformations~\cite{ozaki_2012_error_free_transformations, rump_2008_accurate_floating_point_summation} enable precisely this style of computation. Yet the former commonly rely on maintaining copies of the same matrix in different formats while the latter are sensitive to the dynamic range of the inputs. Thus, even when the numerical algorithm is designed to exploit mixed precision effectively, the underlying data representation can remain a source of redundant storage, conversion overhead, or numerical error.

Despite extensive research in mixed-precision algorithms and accelerator-aware numerical libraries \cite{osti_1814677, UTK2020MPIRReport, Fasi2020TensorCoresNumerics}, \textbf{there is currently no general-purpose data representation that enables efficient, data-agnostic access to multiple floating-point precisions without duplication or conversion overhead.} Existing approaches primarily optimize the arithmetic performed at different precisions; they do not address the more fundamental question of how a single matrix representation can expose multiple computational precision views efficiently. Addressing this representation-level gap is essential for fully realizing the performance potential of modern accelerator hardware.

\subsection{Contributions}
In this work, we introduce \textbf{TOMOS}\footnote{TOMOS comes from the Greek word $\tau o \mu o \varsigma$ (tómos), meaning a ``cut'' or ``partition.'' The name emphasizes our partitioning of each floating-point value into components, enabling mixed-precision computation without duplication.}, a composite multiprecision format and framework that enables simultaneous access to matrices in $32$-bit and $16$-bit floating-point precisions without requiring explicit copies.

The central idea is to partition each $32$-bit floating-point entry into two $16$-bit words, stored in two separate contiguous arrays. One word provides the primary low-precision representation, while the second stores complementary information needed to reconstruct the original $32$-bit value exactly. Precision-specific access is exposed through specialized \texttt{BLAS} kernels that reconstruct values locally, tile-by-tile or register-by-register, during computation.

Our main contributions are:

\begin{enumerate}
    \item \textbf{A composite-precision matrix representation.}
    We introduce FP32/FP16 and FP32/BF16 layouts that expose multiple computational precision views from a single FP32-sized physical matrix representation. Half-precision operations access only the primary $16$-bit words, while single-precision operations combine the primary and complementary words to recover the original FP32 values exactly. The total storage cost therefore matches that of the highest represented precision.

    \item \textbf{Exact reconstruction with numerically useful low-precision encodings.}
    We design an encoding scheme that incorporates rounding information during low-precision conversion while retaining sufficient information to reverse the rounding and reconstruct the original FP32 value exactly. Values outside the normal FP16 range are handled through dedicated encodings that embed exponent offsets into the $16$-bit words. Matrix metadata can then be used to select how these special encodings are decoded, enabling kernels to trade accuracy for performance while preserving the information required for exact FP32 reconstruction.
    
    \item \textbf{Communication- and compute-efficient multiprecision \texttt{BLAS}.}
    We develop specialized \texttt{GEMM} and \texttt{GEMV} kernels that operate directly on the composite representation. Data movement scales with the compute precision: low-precision kernels transfer only the primary $16$-bit words, while higher-precision kernels reconstruct FP32 values locally during computation. The resulting implementations exploit accelerator hardware, including \texttt{tensorcores}, while avoiding explicit matrix duplication and full-format conversion.

    \item \textbf{An end-to-end mixed-precision linear algebra demonstration.}
    We integrate TOMOS into a mixed-precision least-squares solver and show that the representation provides practical performance and memory benefits beyond standalone kernels. On NVIDIA A100 GPUs, TOMOS-based kernels achieve throughput comparable to vendor implementations, while the mixed-precision solver achieves speedups of up to $1.6\times$ over a single-precision implementation without redundant matrix allocations.
\end{enumerate}

An important numerical consequence of the design is that TOMOS
sacrifices one mantissa bit in the primary FP16 encoding to store
control information. Our experiments show, however, that this does
not imply a corresponding loss in numerical quality: nine mantissa
bits combined with rounding information produce lower error in FP16
\texttt{GEMM} and \texttt{GEMV} than ten mantissa bits obtained by
direct truncation. The representation is also extensible. Special encodings can be interpreted differently depending on the target kernel and matrix metadata, and the same design principles can be generalized to additional precision combinations or more than two precision levels.

\textbf{In summary}, TOMOS treats mixed precision not only as an arithmetic
problem, but also as a data-representation problem. It provides a single, data-agnostic matrix layout that supports multiple compute precisions while preserving the storage cost of the highest represented precision and avoiding explicit data duplication.

\section{Background}

\subsection{Notation}
Notation consistent with the IEEE 754 floating-point standard \cite{ieee754_2019} is used throughout this work. A floating-point format that uses $k$ bits is denoted as FP$k$. Common supported types include FP64, FP32, FP16, FP8 and BF16. Floating-point numbers consist of the fields: sign $S$, fractional part (or mantissa) $F$, and exponent $E$, while an exponent bias $B$ is used to simplify comparisons in hardware. Together, these represent either the normalized number $(-1)^{S} \times (1 + F) \times 2^{E - B}$, when $E \neq 0$, or the subnormal number $(-1)^{S} \times F \times 2^{1 - B}$.

Subnormals represent a region of values below the smallest representable normalized exponent by shifting the fractional part to the right. The remaining mantissa bits provide an approximation to numbers in this range. When mantissa information $F$ is lost due to shifting, the representation underflows. If a value exceeds the upper limit of the representable range, an overflow occurs.  The ``not a number'' (NaN) value is used to identify operation-outcomes that do not correspond to a number, e.g. a division by $0$. Table \ref{tab:fp-special-values} describes those encodings.
\begin{table}[!h]
    \centering
    \begin{tabular}{|c|c|c|c|}
       \hline
       Encoding     & $S$    & $E$        & $F$ \\
       \hline
       $\pm \infty$ & $\pm1$ & $E_{\max}$ & $0$ \\
       NaN          & $\pm1$ & $E_{\max}$ & $\neq 0$ \\
       normalized   & $\pm1$ & $\neq 0$   & any   \\
       subnormal    & $\pm1$ & $0$        & any   \\
       \hline
    \end{tabular}
     \caption{Special values of common floating-point systems.}
    \label{tab:fp-special-values}
\end{table}

During conversions, the difference between the source and target exponent-bias is added to the exponent of the input value. We refer to this term as \textbf{\texttt{bias-correction}}; conversion from FP32 to FP16 demands a bias correction of $112$, while $-112$ covers the opposite direction. FP32 and BF16 have the same bias. The notation FP32 / FP16 and FP32 / BF16 provide access to both listed precisions.

\subsection{Related Work}
\label{sec:related-work}

The Customized Precision Mantissa-Segmenting (CPMS) format introduced in
\cite{grutzmacher_2020_pagerank} partitions FP64 values into multiple
segments, storing them either in separate arrays or interleaved layouts
depending on the target kernel and hardware architecture. This design
enables dynamic adjustment of transfer volume by varying the number of
segments loaded into local buffers through efficient encode/decode
routines. The resulting representations retain 11 exponent bits while
allowing a variable number of mantissa bits. In contrast, TOMOS is
designed to expose distinct native compute-precision views, such as FP16
and FP32, from a single FP32-sized representation, rather than varying
the number of mantissa segments used to approximate a higher-precision
value.

Numerical compression techniques
\cite{Lindstrom2006_fpzip, Di2011_SZ, Fox2024_statZFP} reduce memory
requirements by compressing floating-point data and are particularly
relevant when storage and data-movement costs become dominant.
Applications range from large-scale scientific data analysis to deep
neural network training. These methods primarily target storage
reduction, whereas TOMOS preserves the storage footprint of the highest
represented precision and instead reorganizes that storage to enable
direct access at multiple compute precisions.

A different strategy leverages reduced-precision instructions to emulate
higher-bitwidth arithmetic. Such approaches aim at improving the
performance of numerical kernels, including \texttt{GEMM}, while
maintaining high numerical accuracy. Representative examples include APE
\cite{ma_ape_2022} and MixPert \cite{lin_mixpert_2024}, which emulate
32-bit numerics using lower-precision instructions. TOMOS addresses the
complementary problem: rather than emulating one arithmetic precision
using another, it provides multiple precision views of the same stored
data.

High-precision emulation via the shared-exponent mantissa-splitting Ozaki
scheme \cite{mukunoki_2020_dgemm_using_tensor_cores,
ootomo_2024_dgemm_on_integer} relies on an error-free transformation
\cite{ozaki_2012_error_free_transformations} to obtain throughput. The
method first normalizes matrix entries and separates their mantissa bits
into multiple word matrices. The original matrix can then be written as
$
\ma = \ma_1 + \ma_2 + \dots + \ma_k.
$
When applied to matrix multiplication with operands
$\ma = \sum_{i=1}^{k_1}\ma_i$ and
$\mb = \sum_{j=1}^{k_2}\mb_j$, the product is expressed as
$
\sum_{i = 1}^{k_1} \sum_{j = 1}^{k_2} \ma_i \mb_j.
$
Since the terms $\ma_i$ and $\mb_j$ contain subsets of the mantissa bits
of $\ma$ and $\mb$, the products $\ma_i \mb_j$ can be evaluated
efficiently using integer multiplication combined with high-precision
floating-point accumulation. TOMOS also partitions floating-point values,
but for a different purpose: its components form a storage
representation from which either a low-precision value or the original
high-precision value can be accessed directly, without summing multiple
partial products to emulate higher precision.

Finally, exponent-sharing techniques, such as block floating-point
formats \cite{hbfp_2018_drumond, fast_dnn_training_zhang_2022,
adaptive_fp_yeh_2022}, simplify floating-point arithmetic by grouping
values with the same exponent. These formats reduce the computational
complexity of floating-point arithmetic and are widely used in numerical
computations. Unlike block floating-point formats, TOMOS does not rely on
sharing exponents across groups of values and preserves exact
reconstruction of each original FP32 entry.

Overall, prior work has explored precision reduction through
segmentation, compression, arithmetic emulation, and shared-exponent
representations. TOMOS differs in its objective: it is a data
representation for exposing multiple computational precision views of
the same matrix from a single physical layout. Its design preserves exact
recovery of the original FP32 data while allowing low-precision kernels
to access only the subset of stored information required for computation.

\section{TOMOS: Composite Precision Representations}
In this section we present the core design  behind the composite-precision representations and dense matrix layouts in \textbf{TOMOS}.

A 32-bit floating-point value $f$ is converted to the FP32 / FP16 representation by subdividing its fields into two 16-bit words, the \textit{primary} encoding $p$ and \textit{complementary} encoding $c$.
The former is an approximation to the FP16 version of $f$ that has \textit{five} exponent and \textit{nine} mantissa bits (one less than its FP16 encoding); $c$ is a 16-bit residual word which is not a floating-point number in itself, but combined with $p$ reconstructs  the original FP32 value $f$ exactly.

\begin{figure}[!htbp]
   \centering
   \includegraphics[width=\linewidth]{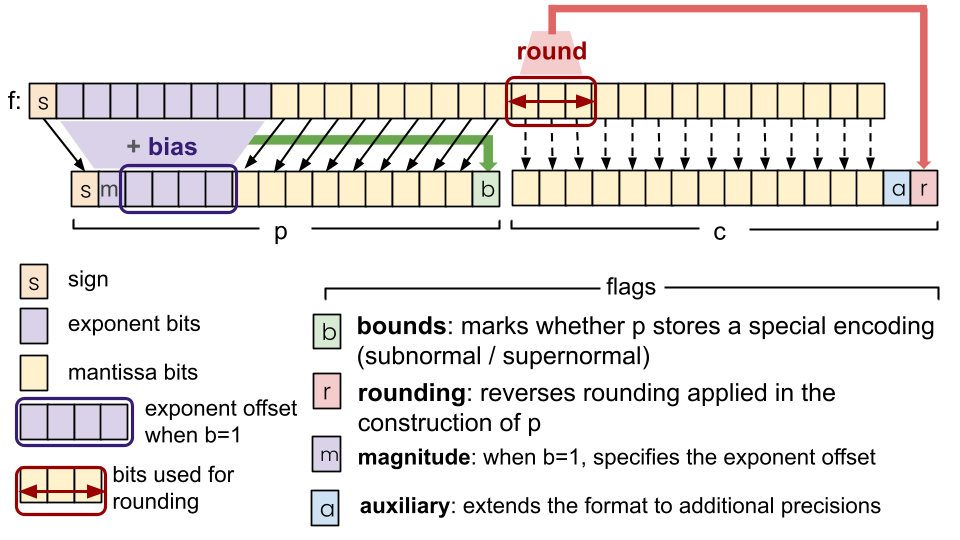}
   \caption{Encoding of FP32 value $f$ into words $p$ and $c$ of the FP32 / FP16 representation; $S$ is the sign bit, $M$ is the 23-bit mantissa (yellow), and $E$ is the 8-bit exponent (purple).  
   The corresponding bias offset is added to the exponent of $f$ to produce the 5-bit FP16 exponent of $p$. 
  Special values set the out-of-bounds flag $b$ to $1$, and $m$ to mark the type of violation. For $b = 1$ and $m=0$ the \textit{four} least significant bits of $p$'s exponent, store the number of right-shifts that map $p$ to an FP16 subnormal; for $b = 1$ and $m = 1$ the exponent holds the offset over the maximum FP16 exponent. A subset of leftover bits are used to apply rounding and $r$ reverse it during FP32 reconstruction. Auxiliary bit $a$ extends the format.}   
   \label{fig:fp_32_16_encoding}
\end{figure}

The FP32 / FP16 encoding scheme is presented in Figure \ref{fig:fp_32_16_encoding}.
In addition to the exponent, mantissa, and sign of $f$, the words
$p$ and $c$ store the control bits $b$ (bounds), $r$ (rounding), and $a$ (auxiliary), while flag $m$ (magnitude) is attached to the exponent when $b=1$. Flag $b$ indicates whether an out-of-bounds violation has occurred during initialization of $p$ from $f$, with $m$ specifying if $p$ holds information related to a subnormal or a supernormal\footnote{A value exceeding the maximum FP16 representation exponent} value. Flag $r$ records the opposite direction $p$ was rounded towards, during low-precision conversion and
 $a$ is an auxiliary bit used to extend the format or pack additional information. 

We refer to the generation of primary and complementary encodings as FP32 / FP16-encode,
and to the reconstruction of $f$ from them as FP32-decode; FP16-decode transforms words that hold special values,  to valid FP16 words.

\subsection{Rounding reversal in FP32 reconstruction}

During the encode step of an FP32 value $f$, its mantissa bits are separated, with the nine most significant bits placed into $p$. While truncation  allows direct reconstruction, it also introduces numerical bias that can quickly dominate computations with low-precision values. By incorporating rounding during the encode step we  overcome this limitation.
Since rounding is non-invertible, direct reconstruction of $f$  is impossible without storing additional information. To address this limitation, the least significant bit of $c$ is reserved as a flag indicating the reverse rounding direction. During the decode stage, this bit is referenced to negate rounding applied in initialization and allow exact reconstruction of $f$.

\subsection{Special Encodings of the FP32 / FP16 Representation}
During the FP32 / FP16-encode step, FP32  inputs may be mapped to FP16 subnormals. To avoid loss of mantissa information, the $9$ most significant bits of the FP32 mantissa field are stored in $p$ (after rounding is applied). The control bits $b$ and $m$ are set to $1$ and $0$ respectively, and the $4$ least significant bits of exponent store the count of right-shifts that convert the stored mantissa to a valid FP16 subnormal. 

Instead of mapping overflows to $\pm$Inf, TOMOS uses the $4$ least-significant exponent bits in a similar way, storing the offset to the maximum FP16 exponent. Both $b$ and $m$ are set to $1$ to mark this state. This  \textit{supernormal} representation can extend the exponent to capture  ``near misses'' during FP32 / FP16-encode. 

Interpretation of special encodings is up to the decode function. An  implementation may  reconstruct subnormals, or map them to $0$ to minimize generated instructions. Similarly, a value within the extended supernormal range can be decoded to a format with $8$ exponent bits to prevent information loss. Effective kernel dispatch takes advantage of matrix metadata collected during initialization. 

\subsection{Decode Device-Function Optimizations}
Efficient encode / decode implementations are essential for computing with data stored in the  FP32 / FP16 representation. We primarily focus on optimizations that can be applied to decode, as it is  integral to BLAS kernels. Encode on the other hand is called once during initialization, its cost amortized over the total runtime. While FP32-decode and FP16-decode are not identical, optimizations target the same underlying operation: unpacking the information attached to the exponent of the primary encoding. For this reason we refer to decode-optimizations  without specifying the output-precision.

We consider   two types of decode implementations: \textbf{conditional}, and \textbf{branchless}. The former follows closely Algorithms 2-3 relying on conditional statements to select an execution path. The disadvantage of this approach is that it causes thread divergence\footnote{Thread divergence occurs when threads within the same warp execute different instructions. Because every thread in a warp operates in a lock-step fashion, different execution paths serialize thread computations.} within warps.  The number of affected warps depends on the number and distribution of special encodings.
Branchless implementations follows every path to avoid thread divergence, and selects the output through logical instructions. Although this strategy requires more resources, and generates a larger number of assembly instructions, it is independent to the statistics of matrix-elements. Furthermore, \textbf{fast-decode} functions map subnormals to zero.

\begin{algorithm}[htbp]
\caption{[$p$, $c$] = \textbf{\texttt{FP32/FP16-encode}}($f$)
}
\begin{algorithmic}[1]
    \State [mantissa-high, mantissa-low, r-direction] $\leftarrow $ \textbf{\texttt{round}}(f, $9$); 
    \State exponent $\leftarrow$ f.exponent + bias-correction
    \If{(exponent $\geq E_{\max}^{\text{FP16}}$)}
        \State b $\leftarrow$ 1; m $\leftarrow$ 1;  \begin{color}{dark-green}// encode as overflow, store offset \end{color}
        \State exponent $\leftarrow$ overflow-offset(exponent)
    \ElsIf{(exponent $\leq E_{\min}^{\text{FP16}}$)}
        \State b $\leftarrow$ 1; m $\leftarrow$ 0; \begin{color}{dark-green}// encode as underflow and store offset\end{color}
        \State exponent $\leftarrow$ underflow-offset(exponent)
    \Else
        \State b $\leftarrow 0$;  \begin{color}{dark-green}// encode as normal value\end{color}
        \State exponent $\leftarrow$ truncate-to-5bits(exponent)
    \EndIf
    \State p.b $\leftarrow$ b; p.m $\leftarrow$ m; c.r $\leftarrow$ r-direction
    \State p.mantissa $\leftarrow$ mantissa-high; c.mantissa $\leftarrow$ mantissa-low 
    \State p.exponent $\leftarrow$  exponent
    \end{algorithmic}
\label{alg:unpack-to-fp16_0}
\end{algorithm}

\begin{algorithm}[htbp]
\caption{[$f$] = \textbf{\texttt{FP32-decode}}($p$, $c$)
}
\begin{algorithmic}[1]
    \State x $\leftarrow$ p
    \If{c.r == 0}
        \State x$\leftarrow$ round-down(p);
    \EndIf
    \State mantissa $\leftarrow$ [x.mantissa, c.mantissa]
    \If{p.b == 0}
    \State exponent $\leftarrow$ x.exponent - bias-correction
    \Else  \begin{color}{dark-green} \ \ // unpack offset in x.exponent \end{color}
        \State shift $\leftarrow$ x.exponent[4]
        \State exponent $\leftarrow$ unpack-offset(x.exponent[3:0], shift)
    \EndIf
    \State f $\leftarrow$ [p.sign, exponent, mantissa]
\end{algorithmic}
\label{alg:pack-to-fp32_0}
\end{algorithm}
\begin{algorithm}[!htbp]
\caption{[$h$] = \textbf{\texttt{FP16-decode}}($p$)
}
\begin{algorithmic}[1]
    \If{p.b == 0}
    \State h $\leftarrow p$
    \Else
        \State shift $\leftarrow$ p.m
        \If{shift == 1}
        \State h $\leftarrow \pm$Inf
        \ElsIf{shift == 0}
        \State mantissa $\leftarrow$ p.mantissa $>>$ p.exponent[3:0]
        \State h $\leftarrow$ [p.sign, $0$, mantissa]
        \EndIf
    \EndIf
\end{algorithmic}
\label{alg:pack-to-fp32_0}
\end{algorithm}
\begin{figure}[!htpb]
   \centering
   \includegraphics[width=\linewidth]{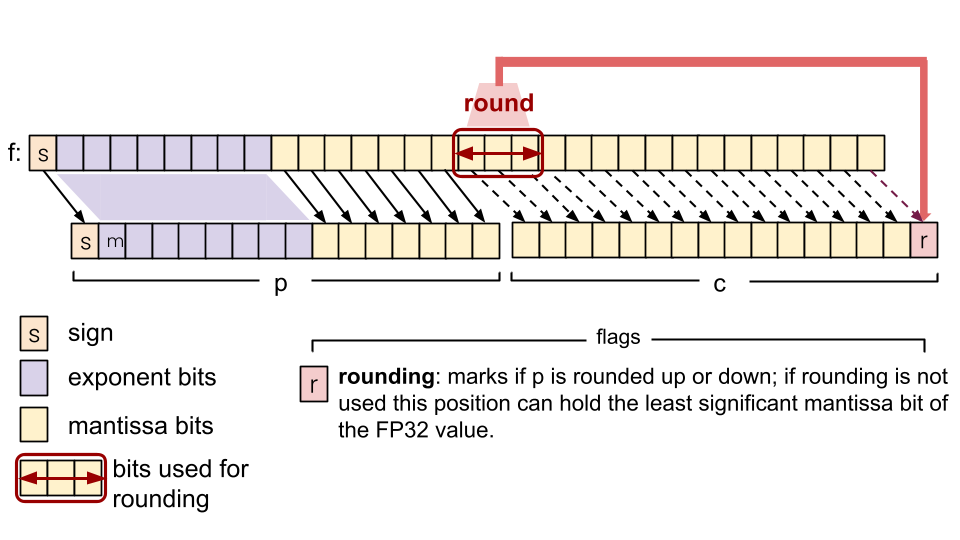}
   \caption{Encoding FP32 value in FP32 / BF16 representation.
   }
   \label{fig:encode_fp32_bf16}
\end{figure}

\subsection{FP32 / BF16 Representation}
The FP32 / BF16 format uses primary encodings with $8$-bit exponent, resulting in the design of Figure \ref{fig:encode_fp32_bf16}. The format enables computations in FP32 and BF16 precisions and  is simpler to optimize for, since no special encodings are required. Rounding during low-precision conversions is optional. If used, the format exchanges the least significant mantissa bit of the FP32 representation to store the rounding flag. Its primary encodings have $2$ fewer mantissa bits  compared those of the FP32 / FP16 layout. 

\subsection{Multiprecision Access to Contiguous Matrix Entries}

Within TOMOS, a BLAS kernel can access both FP32 and FP16 versions of
a matrix by leveraging reads over contiguous storage.
This is accomplished by splitting matrix $\ma$ of FP32 entries into matrices $\ma_p$ and $\ma_c$ that contain the primary and
complementary encodings respectively.
This construction provides direct access to FP16 values through $\ma_p$ and implicit
access to FP32 matrix values by transferring both $\ma_p$ and $\ma_c$
into registers and decoding them. 
Since reconstructing $\ma$ in its entirety  contradicts our main objective of minimizing data-duplication,
FP32 values are assembled in parts, within GEMM and GEMV routines to evaluate a matrix action in the specified precision.

The composite matrix components are accessed blockwise; small tiles of $\ma$ are reconstructed to FP32, to multiply tiles of $\mb$. This approach achieves the same memory footprint and communication cost as native FP32 implementations. The  latency introduced by the \texttt{encode} and \texttt{decode} steps is mitigated in TOMOS by overlapping them with memory transfers and numerical computations.

\section{Multiprecision BLAS on the NVIDIA A100}
\label{sec:multiprecision-blas}

Composite-precision formats alter memory organization, which requires BLAS re-implementation. These kernels have to support efficient element-packing into native 32-bit or 16-bit floating-point types, and interleave computations with data-transfers to mitigate loss of spatial locality. We provide implementations for FP32, FP16 or BF16 compute-precisions,  over data in the storage-types  listed bellow:
\begin{enumerate}
    \item GEMM: FP32, FP16, BF16, FP32 / FP16, FP32 / BF16
    \item GEMV: FP32, FP16, BF16, FP32 / FP16 and their transposed variants 
\end{enumerate}
We also developed baselines in native FP32, FP16 and BF16 precision to use as a starting points, and to identify bottlenecks introduced by the composite matrix-layouts.   

\subsection{GEMM Kernels}
The $32$-bit GEMM kernels, ``TOMOS FP32'', rely on established optimizations. Thread-blocks evaluate an $128 \times 128$ output tile, by multiplying $128 \times 8$ sub-blocks of $A$ with $8 \times 128$ sub-blocks of $B$. This is a well tested strategy for interleaving compute and transfer instructions. A $1$D thread-block topology  with $256$ threads per block is used that accesses the input matrices via synchronous, vectorized loads. The tile-product computation is organized in a four-layer pipeline to maximize throughput.

Incrementing the number of pipeline-levels reduces synchronization but increases shared-memory usage. The partial tile-product is evaluated through a fine-grained $2$-stage process. The output tile is moved to global memory through scalar stores, which is a notable difference with state-of-the-art implementations that rely on coalesced store instructions.

The FP32 / FP16 kernels has access to $A$ in the composite layout, while $B$ is stored in a single contiguous array of 32-bit floats. The decode-cost and added temporary variables required to address elements in   $A_p$ and $A_c$ increase register-contention and limit the number of active tiles.
To overcome this limitation, the best performing FP32 / FP16 kernels use a tiled, two-layer asynchronous pipeline that takes advantage of intra-warp instruction-interleaving by executing FMA instructions over issued asynchronous tile transfers.
As shown in the code section \ref{code:fp32_fp16_gemm}, this strategy cannot  interleave decode steps. While in theory, a three-stage pipeline based on asynchronous loads  accomplishes this task, our benchmarks revealed  consistent slowdowns due to increased shared-memory requirements.

\begin{lstlisting}[caption={Hot-loop structure followed by  32-bit GEMM on the FP32 / FP16 matrix layout}, label={code:fp32_fp16_gemm}]
issue_transfer(next_tile_A, next_tile_B)
compute(current_tile_A, current_tile_B)
wait_for_transfer(next_tile_A)
decode_(next_tile_A)
\end{lstlisting}

One way to overcome this limitation is to decode batches of the $\ma$ into a global-memory workspace.  This alleviates pressure from shared memory and stages the computation in batches, allowing maximum reuse of reconstructed FP32 values. This scheme also enables stream-level parallelism in exchange for additional workspace allocations. Overall, the total memory savings are given by the  formula:
$$100 \times \left(1 - \frac{\text{num\_elems}(A)\cdot 16 + \text{num\_elems(workspace)}\cdot 32}{\text{num\_elems}(A) \cdot (32 + 16)}\right) \%$$ 
As long as the workspace memory holds a small subset of  matrix rows, memory-reduction benefits remain  unaffected.

The FP16 reference \texttt{GEMM}, evaluates output tiles of size $256 \times 128$,   achieving lower occupancy \footnote{Occupancy refers to the ratio of warps that are active per streaming multiprocessor} in exchange for deeper software pipelining. A $4$-stage pipeline, processes tiles using MMA instructions \footnote{Matrix-Multiply-Accumulate instructions: those are PTX instructions (intermediate representation that sits between CUDA and the hardware), that dispatch compute-instructions to tensorcore units}. The ``TOMOS BF16'' method follows the same logic, replacing compute-instructions to match the BF16 operands.

The FP32 / FP16 kernel accesses only $A_p$ and is similar to ``TOMOS FP16''; when all entries of $A$ are within the FP16 representation range, $A_p$ holds valid FP16 values and packed subnormals.
Branchless and fast-decode, combined with workspace-based methods achieve high throughput.

\subsection{GEMV Kernels}
Due to the large number of GEMV implementations, we describe the strategies applied for large-scale inputs.

The ``TOMOS FP32'' reference GEMV kernel does not use any shared memory for caching values of $A$ or $x$. Every thread is mapped to a row of the output vector, and is completely responsible for evaluating the dot product. Vectorized loads are used for data transfers, and FMA instructions perform the numerical computations. Stores  are not coalesced, since every thread writes  a single element.  The best-performing FP32 / FP16 kernel uses a similar strategy  to ``TOMOS FP32'';  elements of $A_p$ and $A_c$ are accessed in groups, decoded to FP32, with the dot product updated via FMA instructions. 

The FP16 GEMV kernel is structured similarly to ``TOMOS FP32'' GEMV, with \_\_half2 instructions used instead \footnote{\_\_half2 instructions correspond to SIMD FMA instructions.}. Partial sums are updated using \texttt{atomicAdd}.

Branchless decode implementations are crucial for achieving high throughput in either FP32 or FP16 compute; workspace-based techniques do not provide an advantage for GEMV, as those are memory-bound.

\section{Evaluation Methodology}
We evaluate the proposed representations in TOMOS as matrix layouts for enabling mixed-precision computations. We begin by estimating the performance and numerical accuracy of standalone matrix kernels. After establishing that efficient \texttt{BLAS} can be built on top of the proposed representations, we use those to design a mixed-precision least-squares solver and highlight end-to-end performance advantages and effective memory-usage reduction within a numerical method.  
We also investigate the role of rounding and subnormal representations in computations with the low-precision encodings.  

\subsection{Standalone Kernel Evaluation - Performance}

We first ask whether the additional encoding and decoding required by
TOMOS can be hidden sufficiently to preserve the performance of native
floating-point representations. We evaluate this question separately
for compute-bound \texttt{GEMM} and memory-bound \texttt{GEMV}
operations, comparing TOMOS against our native-precision baselines,
cuBLAS, and CUTLASS.

\subsubsection{Notation} 
Labels ``(fast)'' decode methods. Conditional decode is used for 32-bit compute GEMM, and branchless decode used by 16-bit GEMM and GEMV kernels.
Methods labeled as ``(workspace)'' use additional scratch memory; the ``(truncated)'' variants skip rounding during construction of low-precision encodings.

\subsubsection{Test cases}
We consider two instances of matrix-matrix multiplication build on top of the general formula $C \leftarrow C + AB$ for $m \times k$ matrix $\ma$, and $k \times n$ $\mb$. First we target multiplication of square matrices where $m = n = k$, and then focus on instances where $m$ is equal to $n$ and $k \ll m$.
We report performance in TFLOP/s for increasing number of matrix-dimensions. For rectangular operands, $m$ and $n$ are fixed to $m = n$, and throughput reported for increasing values of internal-dimension $k$. The 32-bit GEMM uses conditional decode; despite this, we  show that workspace-based methods do not increase runtime depending on the subnormal statistics. The 16-bit GEMM kernels use branchless decode thus performance is independent to subnormal statistics. Similar experiments are performed for GEMV. Both 32-bit and 16-bit methods use branchless decode. Transposed GEMV kernels achieve similar performance, thus their plots are omitted. 

\subsubsection{Conventions}
In order to achieve fair comparisons we first implement native-precision GEMM and GEMV kernels to compare against. We follow this methodology to comment on the effects that the matrix layouts have on a particular computation. 
Furthermore, performance comparisons mostly refer to asymptotic behavior except if mentioned otherwise, since large-scale matrices correspond to realistic instances where a competitive advantage is provided. For this reason, our optimizations mostly target large-scale inputs.

\subsection{Standalone Kernel Evaluation - Numerical Accuracy}

We next ask whether obtaining low-precision values from the composite
representation introduces a meaningful numerical penalty relative to
native floating-point storage. This question is particularly important
for FP32/FP16, since the primary encoding uses nine mantissa bits rather
than the ten available in standard FP16.

\subsubsection{Notation}
Error plots quantify the numerical accuracy of the developed kernels. We report the relative Frobenius error  obtained through the following procedure. First, input matrices $\ma$ and $\mb$ are initialized in high precision (FP64 or FP32), then demoted  to FP32 or FP16 respectively;  $\ma$ is stored in a composite format while $\mb$ is stored in a native floating-point representation. The experiments compute $\mc \leftarrow \text{demote}(\ma) \cdot \text{demote}(\mb)$ and  relative error as $\frac{\|\text{promote}(\mc) - \ma \mb\|_F}{\|\ma \mb \|_F}$.  

\subsubsection{Test cases}
 Numerical comparisons are performed for FP32, FP16 and BF16 inputs to \texttt{GEMM} and \texttt{GEMV} operations, over randomized matrices in the following settings: (a) elements of square $n \times n$ matrices $A$ and $B$ sampled uniformly from $\begin{bmatrix}-1, 1 \end{bmatrix}$, with error reported for  increasing values of $n$; (b) the number of rows and columns is fixed to $n = 32768$, generating $A$$\sim$$\text{Uniform}(-M , M)$ and $B$$\sim$$\text{Uniform}(-1/\sqrt{n}, 1/\sqrt{n})$ to report relative error for scaling factor $M$; (c) for the same setting as (b) we compare the error obtained using the ``TOMOS FP16'' GEMM on inputs  converted from FP32 via different rounding schemes to determine the effect of rounding in computations with TOMOS's low-precision encodings.

\subsection{Mixed-Precision Linear Regression Solvers}
\label{sec:lsqr}

A mixed-precision LSQR solver is used to demonstrate the end-to-end
performance and memory benefits of TOMOS within a numerical method.
For the overdetermined least-squares problem
$\vx^{*} = \arg\min_{\vx\in\mathbb{R}^n}\|\vb-\ma\vx\|_2^2$,
we use the LSQR iterative solver introduced in
\cite{householder_unitary_triangulation, givens_rotations_1958,
golub_kahan_svd, golub_reinsch_svd, hestenes_cg, paige1982lsqr}.
We pair LSQR with a CountSketch-based randomized preconditioner
\cite{charikar_finding_frequent_items,
cormode_countsketch/j.jalgor.2003.12.001,
clarkson_low_rank_approximation} to improve convergence, following
the well-established randomized least-squares paradigm
\cite{drineas_sampling_algorithms_2006,sarlos_random_projections,
avron2010blendenpik,drineas_a_faster_approximation,
clarkson_low_rank_approximation}.

We perform experiments with synthetic problems that challenge LSQR's convergence. We consider matrices with $10^6$ rows, a varying number of columns and condition number of $1000$. Their singular values follow a geometric distribution with the largest being equal to one (see Table \ref{tab:matrices}).
\begin{table}[]
    \centering
    \begin{tabular}{|c|c|c|c|c|r|}
    \hline
    \textbf{name} & \textbf{m} & \textbf{n} & \textbf{cond} & $\boldsymbol{\sigma}_{\boldsymbol{\min}}$ & \textbf{underflow} $\%$ \\
    \hline
    matrix-1 & $10^6$ & $10^3$ & $1000$ & 1 & $0.043\%$ \\
    matrix-0 & $10^6$ & $10^3$ & $1000$ & 1e-3 & $18\%$ \\
    \hline
    \end{tabular}
    \caption{Synthetic test matrices used in LSQR experiments. Singular values follow geometric distribution, $\sigma_{\min} = 1$.
    }
    \label{tab:matrices}
\end{table}

LSQR supports a mixed-precision scheme that includes the following  compute-types: refine\_t for residual computation; value\_t for BLAS1; mv\_t: for BLAS2; precond\_t for preconditioner application. 
Single-precision implementations set those to FP32, while the mixed-precision implementation perform the matrix-vector products internal to the method in FP16 precision, with the residual evaluated once per restart cycle of $5$ iterations in FP32. The iterations terminate once relative residual norm error $\leq 10^{-7}$ has been achieved. 

\subsection{Reproducibility}

Correctness and performance tests ran on the Gilbreth cluster, at the Rosen Center of Advanced Computing at Purdue University. Experiments were performed exclusively on nodes equipped with A100 Nvidia GPUs (40Gb), AMD EPYC 7763 64-core processors and 512Gb of DDR4 memory per node. The systems runs Rocky Linux $9$, and supports CUDA 12.6, GCC 11.4 and CMake version 3.13.11. Profiling data were gathered through Nsight Compute v2025.4.0.0.

\section{Numerical Experiments}

\subsection{Standalone Kernel Evaluation - Performance}

\subsubsection{GEMM: Can composite storage match native throughput}
\begin{figure*}[!htbp]
    \centering
    \begin{subfigure}[t]{0.49\textwidth}
        \includegraphics[width=0.85\linewidth]{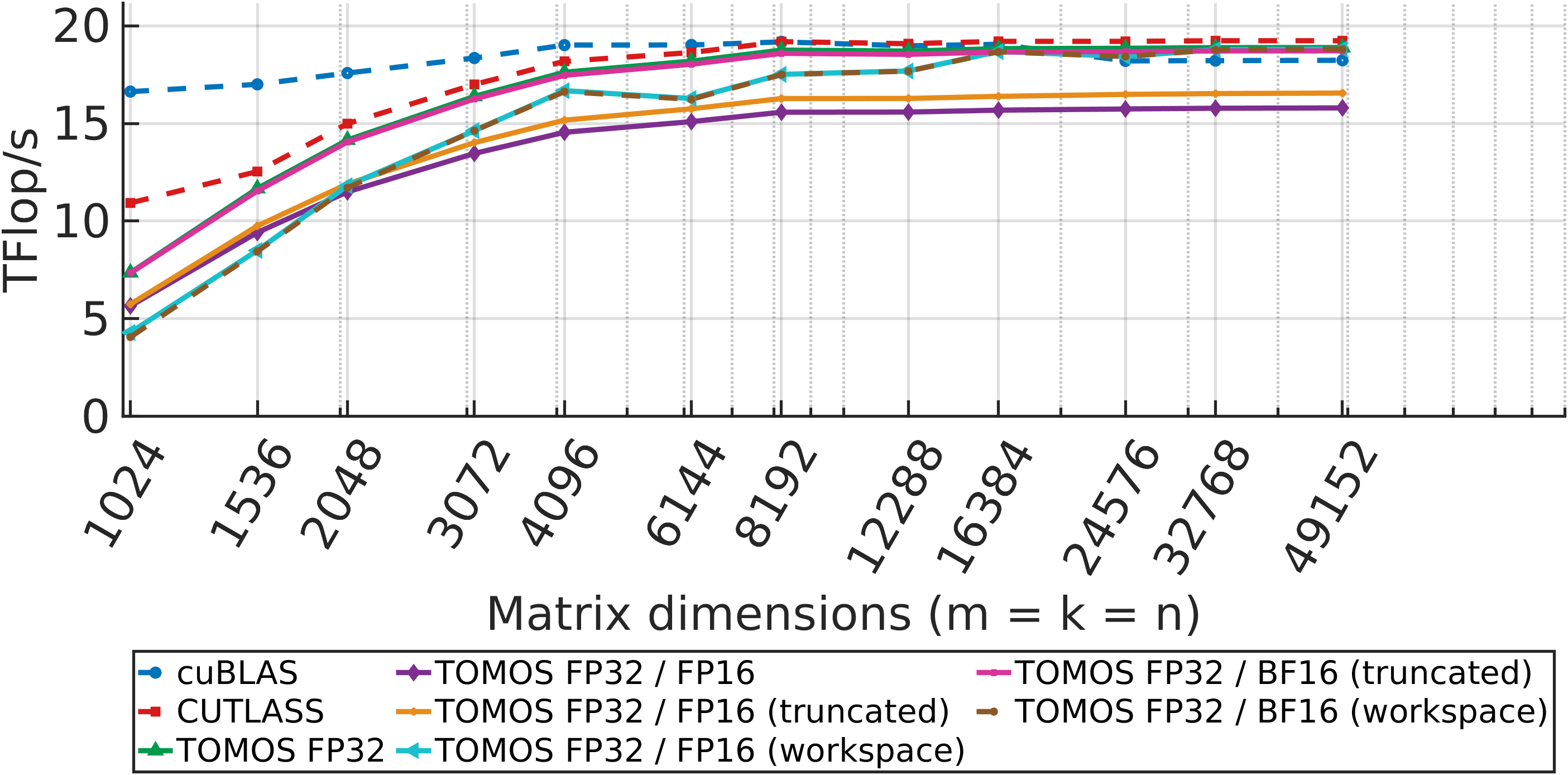}
    \end{subfigure}
    \begin{subfigure}[t]{0.49\textwidth}
        \includegraphics[width=0.85\linewidth]{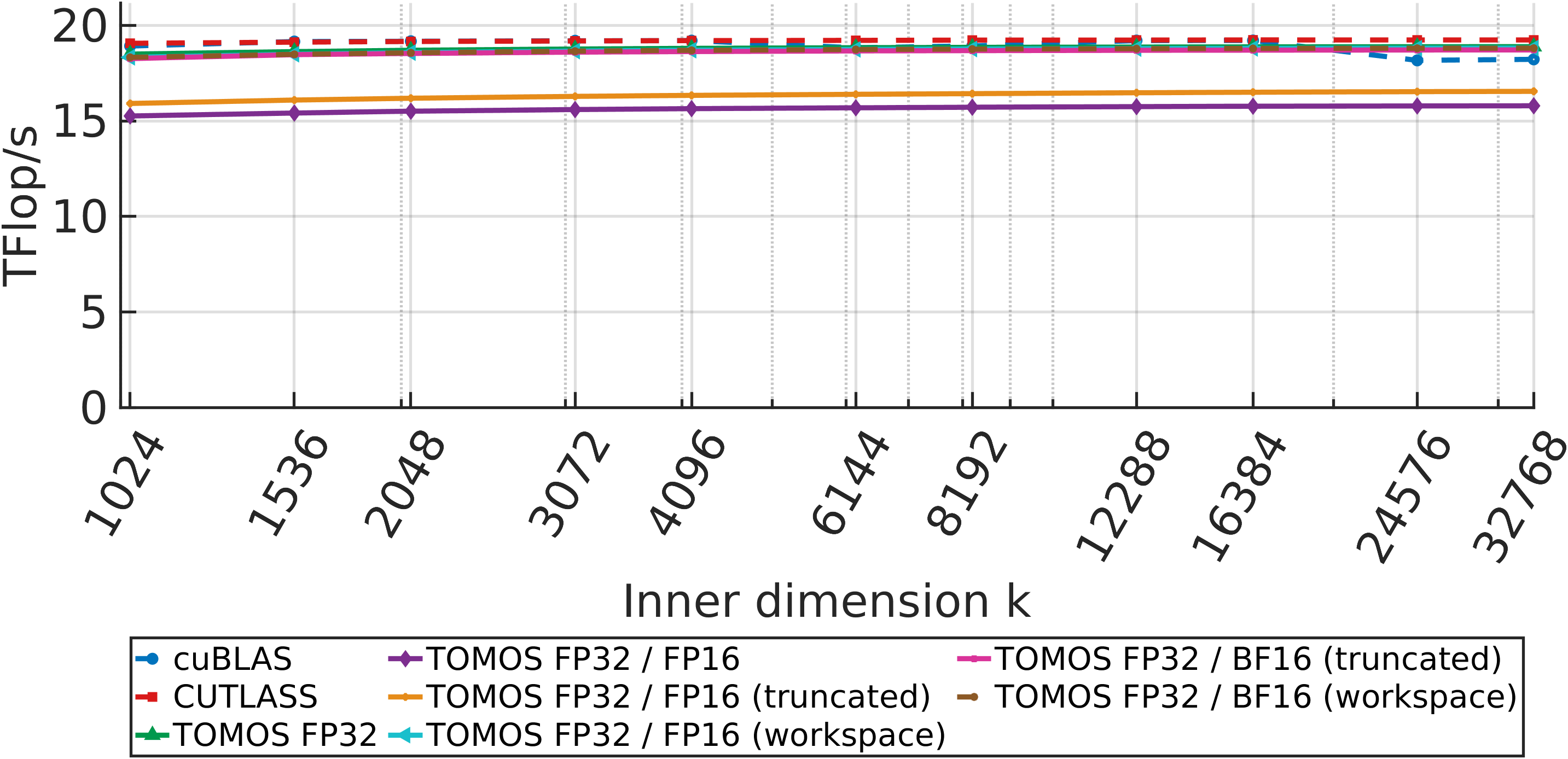}
    \end{subfigure}
        \caption{Throughput for 32-bit kernels. Left: square matrix-matrix multiplication for increasing number of dimensions $m = n = k$; Right: tall-and-thin times a short-and-wide matrix for $m = n = 49000$, for increasing internal dimension $k$.}
        \label{gemm32}
\end{figure*}

Figure \ref{gemm32} displays the throughput achieved by the 32-bit GEMM methods. Results for square matrices are depicted on the left and for rectangular matrices on the right. The reference ``TOMOS FP32'' matches closely cuBLAS performance for $m\geq 8192$ and CUTLASS for all $m$, while FP32 / FP16 kernels achieve a $\sim15\%$ performance drop in comparison.  
The workspace-based construction overcomes those slowdowns. with memory-savings decreasing slightly for $m = n = 49152$, from $33\%$ to $31 \%$ in ``FP32 / FP16 (workspace)''. Rounding-reversal is shown to not impact performance.
The ``TOMOS FP32 / BF16'' kernel matches  the reference FP32 performance without the need for external memory, due to low decode-complexity. Table \ref{tab:subnormal-gemm} demonstrates that the workspace-based methods achieve consistently high throughput for increasing number of subnormals in operands.  
\begin{table}[!htbp]
    \centering
    \begin{tabular}{|l|c|c|c|c|c|}
       \hline
       \textbf{subnormal density} & $0\%$  & $10\%$ & $20\%$ & $30\%$ & $40\%$ \\
       \hline
       \textbf{throughput (TFLOP/s)}  & 18.81 & 18.81 & 18.82 & 18.81 & 18.81 \\
       \hline
    \end{tabular}
    \caption{Throughput of 32-bit ``TOMOS FP32 / FP16 (workspace)'' GEMM, for increasing number of subnormal-density in input operands.}
    \label{tab:subnormal-gemm}
\end{table}

\begin{figure*}[!htbp]
    \centering
    \begin{subfigure}[t]{0.49\textwidth}
        \includegraphics[width=0.85\linewidth]{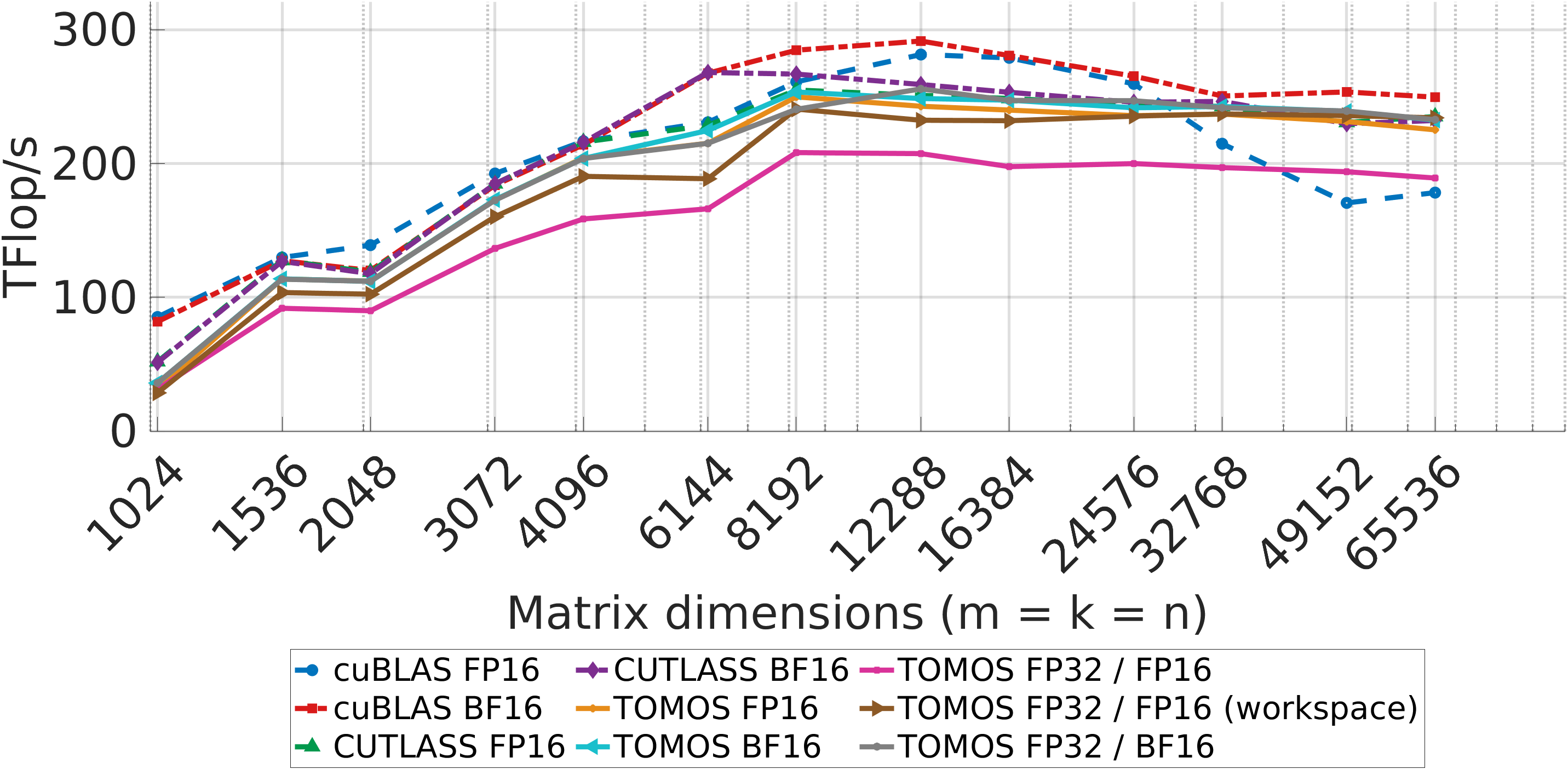}
    \end{subfigure}
    \begin{subfigure}[t]{0.49\textwidth}
        \includegraphics[width=0.85\linewidth]{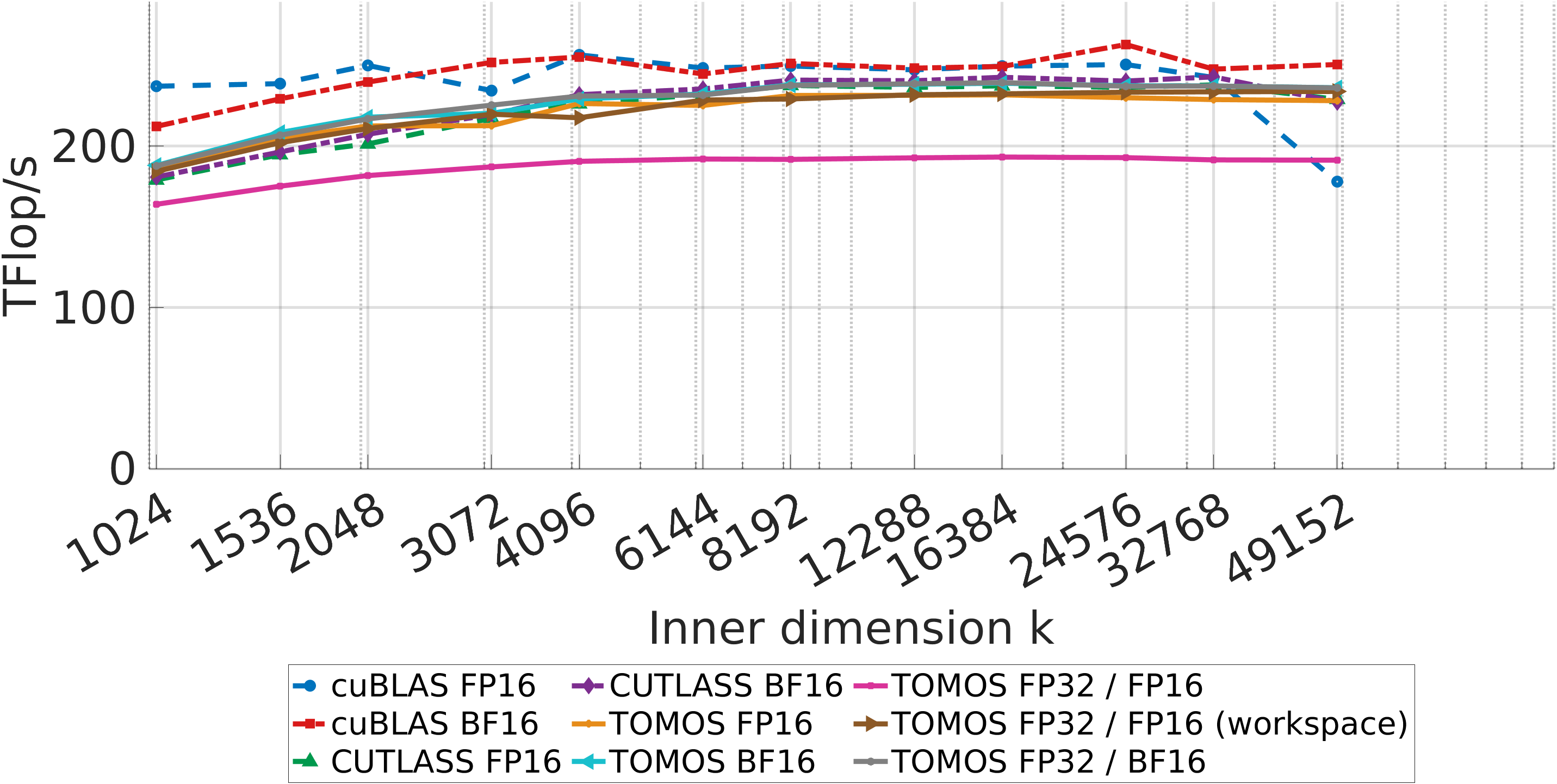}
    \end{subfigure}
        \caption{Throughput for 16-bit kernels. Left: square matrix-matrix multiplication for increasing number of dimensions $m = n = k$; Right: tall-and-thin times a short-and-wide matrix for $m = n = 49000$, for increasing internal dimension $k$.}
        \label{gemm16}
\end{figure*}

In Figure \ref{gemm16} ``TOMOS FP16'' and ``TOMOS BF16'' \texttt{GEMM} achieve throughput that is close to  vendor libraries. The FP32 / FP16, 16-bit  \texttt{GEMM} faces similar performance degradation to its 32-bit counterpart, with the workspace-based strategy used to overcome it. Branchless decode implementations are used, thus subnormal statistics in inputs do not affect performance. The FP32 / BF16 
on the other hand directly dispatches ``TOMOS BF16''.
Those observations are consistent across multiplication of square and rectangular matrices.

These results answer our question affirmatively: for compute-bound matrix multiplication, the cost of decoding the composite representation can be largely hidden, allowing TOMOS to approach native and vendor-library throughput in both 32-bit and 16-bit compute.

\subsubsection{\texttt{GEMV}: What is the cost of decoding in a
memory-bound kernel?}
\begin{figure*}[!htbp]
    \centering
    \begin{subfigure}[t]{0.49\textwidth}
        \includegraphics[width=0.85\linewidth]{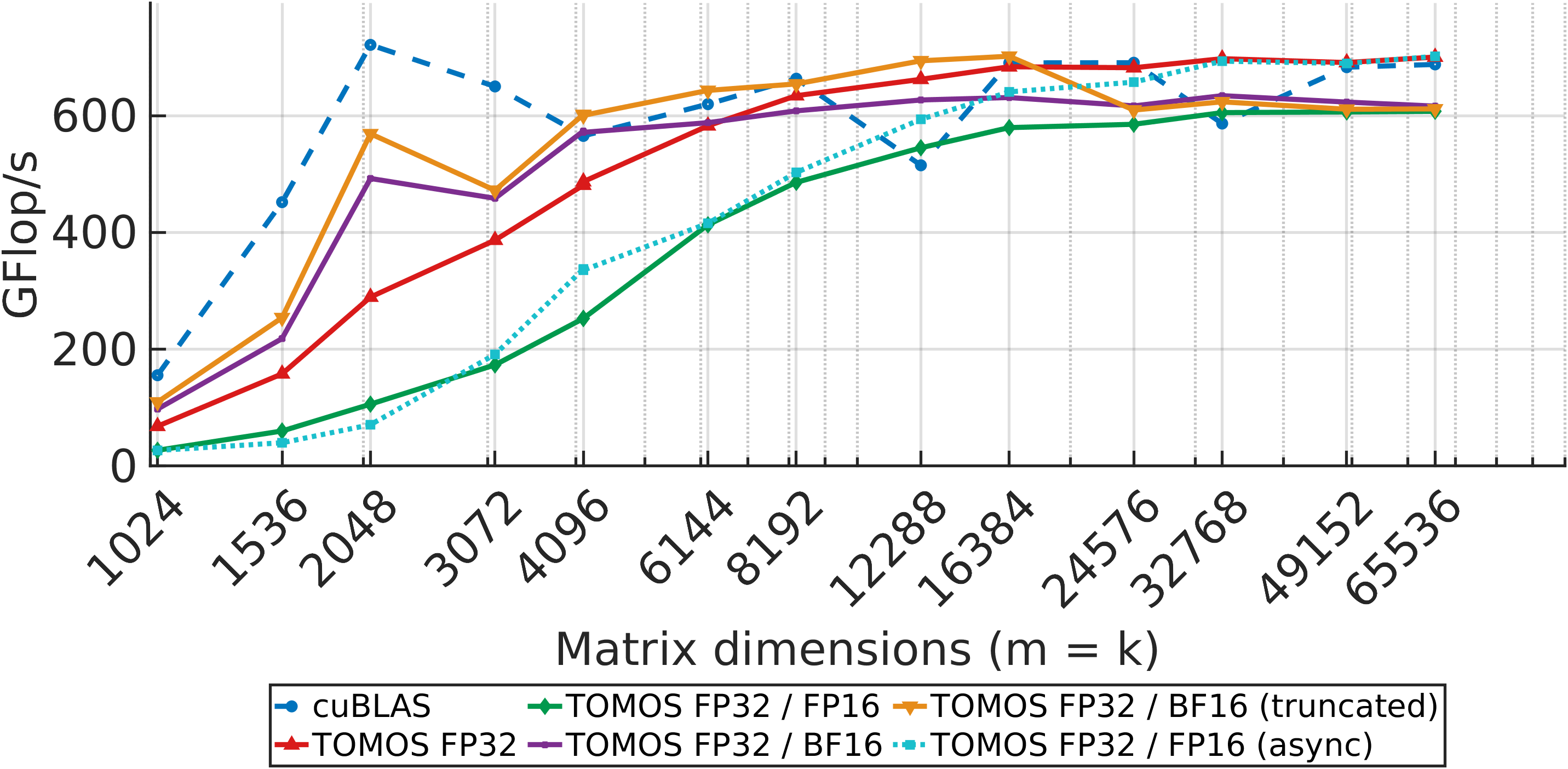}
    \end{subfigure}
    \begin{subfigure}[t]{0.49\textwidth}
        \includegraphics[width=0.85\linewidth]{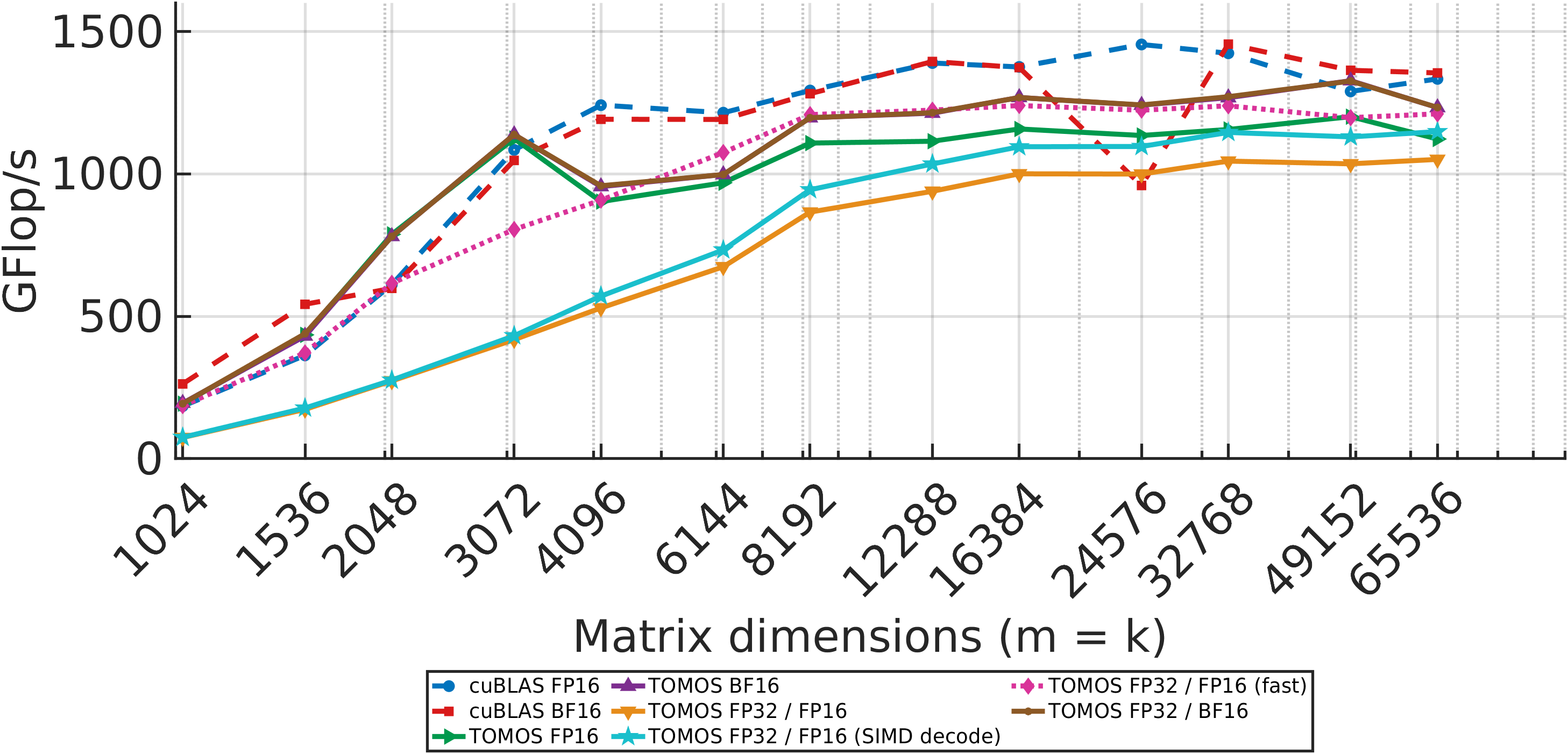}
    \end{subfigure}
        \caption{On the left: throughput for 32-bit \texttt{GEMV} kernels. On the right: throughput for 16-bit \texttt{GEMV} kernels. Both experiments run on square matrices for increasing number of dimensions.}
        \label{gemv32}
\end{figure*}

Matrix-vector multiplication provides a different test because \texttt{GEMV} is memory-bound and offers substantially less computation with which to overlap decoding. Figure~\ref{gemv32} therefore measures how strongly the composite representation affects throughput when decode latency is more exposed.

The left plot depicts the throughput achieved by the 32-bit GEMV implementations; ``TOMOS FP32'' reports comparable performance to cuBLAS and CUTLASS, while ``TOMOS FP32 / FP16'' underperforms. The ``FP32 / FP16 (fast)'' method reduces decode-latency significantly, while FP32 / BF16 is able to match FP32 performance. The throughput achieved by the $16$-bit \texttt{GEMV} kernels is depicted on the right. As in the previous experiments, slowdowns for matrices in the FP32 / FP16 layout occur due to conditional decoding. The fast-decode and branchless decode overcome those slowdowns, while once more the FP32 / BF16 and FP32 / BF16 (truncated) achieve close-to optimal performance.

The results show that decode strategy is considerably more important for \texttt{GEMV} than for \texttt{GEMM}. Branchless and fast decoding recover much of the lost throughput, whereas workspace-based decoding provides little benefit because the operation remains memory-bound.

\subsection{Standalone Kernel Evaluation - Numerical Accuracy}
\subsubsection{Error analysis for scaling matrix-dimensions}

We now compare the numerical error of TOMOS-based \texttt{GEMM} and \texttt{GEMV} with native and vendor implementations as matrix dimension and input magnitude vary. Across the tested randomized inputs, the FP32 TOMOS implementations
exhibit accuracy comparable to the corresponding native implementations. For 16-bit computation, the composite FP32/FP16 representation remains competitive with native FP16 despite using one fewer mantissa bit in
its primary encoding.

\begin{figure*}[!htbp]
    \centering
    \begin{subfigure}[t]{0.49\textwidth}
        \includegraphics[width=0.82\linewidth]{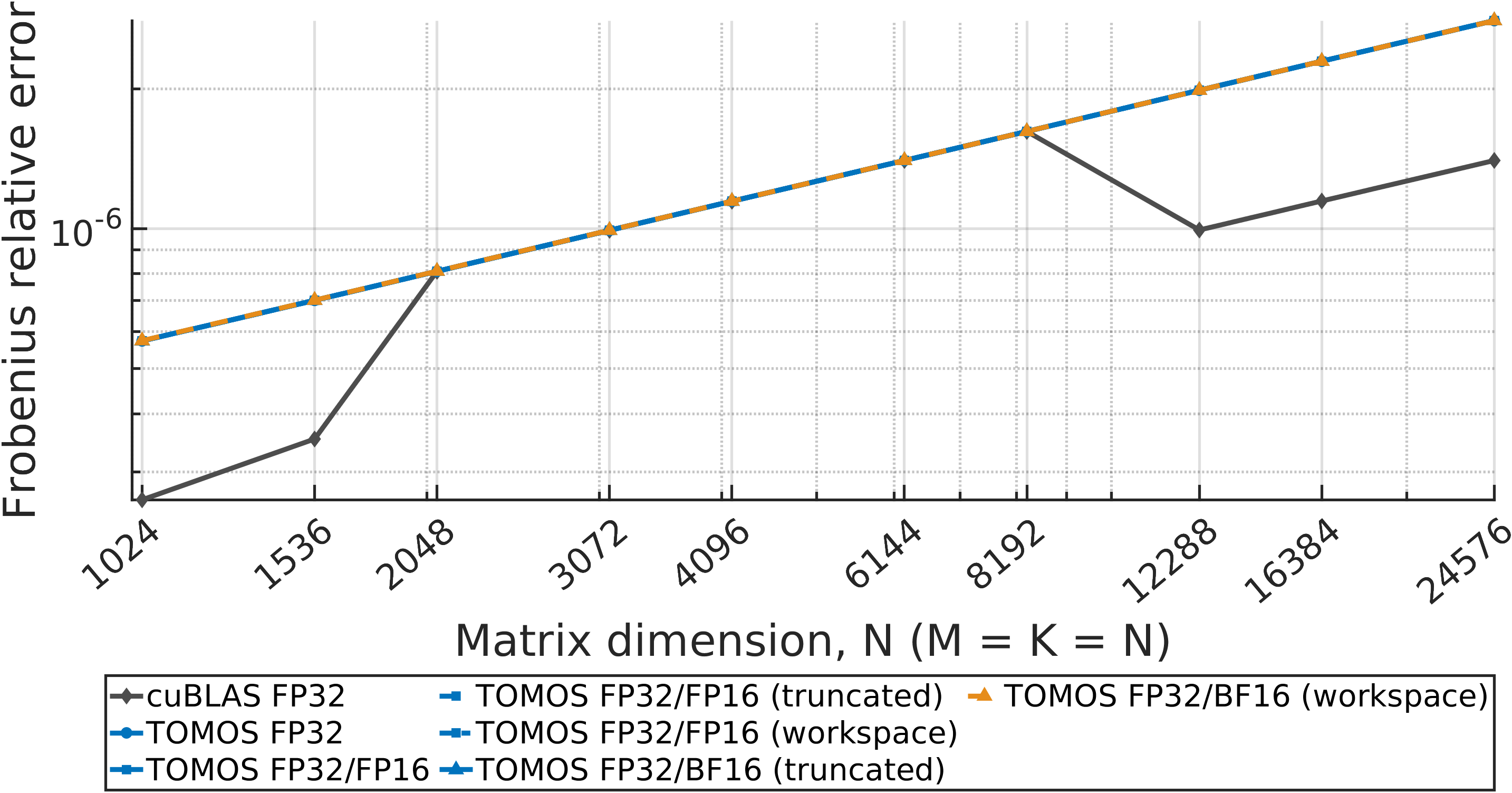}
    \end{subfigure}
    \begin{subfigure}[t]{0.49\textwidth}
        \includegraphics[width=0.82\linewidth]{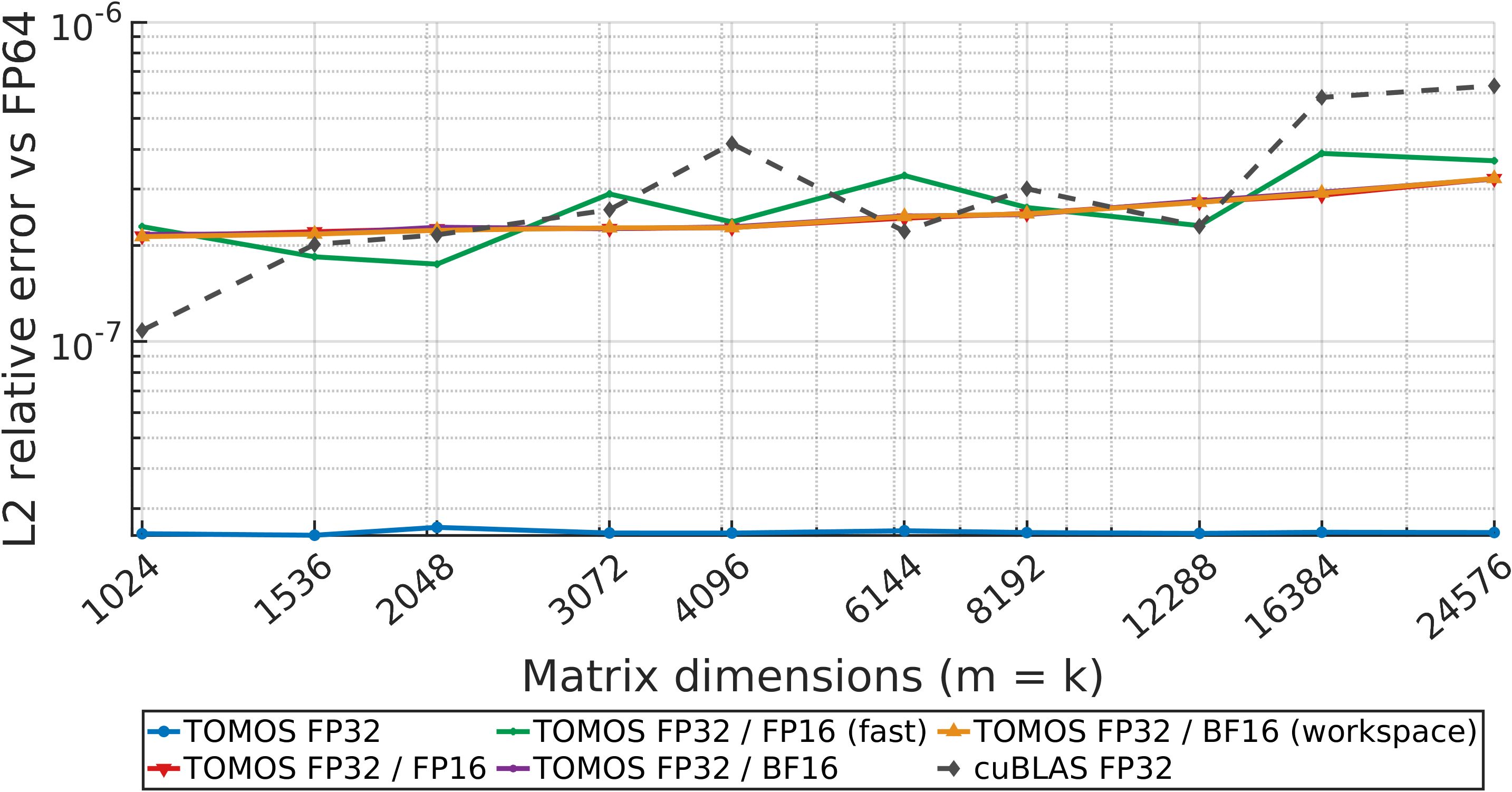}
    \end{subfigure}
        \caption{Relative Frobenius error for 32-bit kernels. Error for \texttt{GEMM} kernels are plotted on the left and \texttt{GEMV} on the right.} 
        \label{error-blas-32}
\end{figure*}

\begin{figure*}[!htbp]
    \centering
    \begin{subfigure}[t]{0.49\textwidth}
        \includegraphics[width=0.85\linewidth]{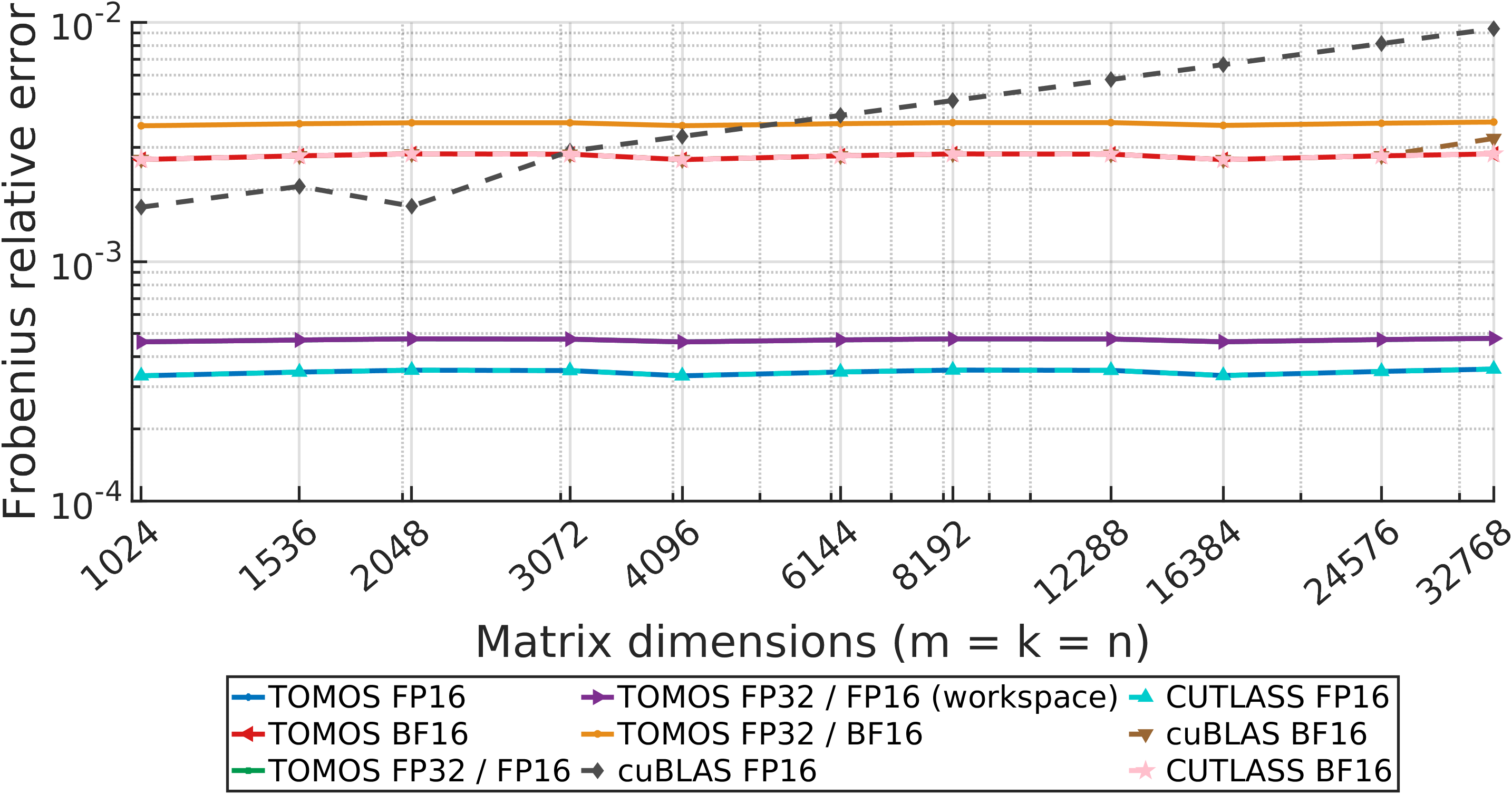}
    \end{subfigure}
    \begin{subfigure}[t]{0.49\textwidth}
        \includegraphics[width=0.85\linewidth]{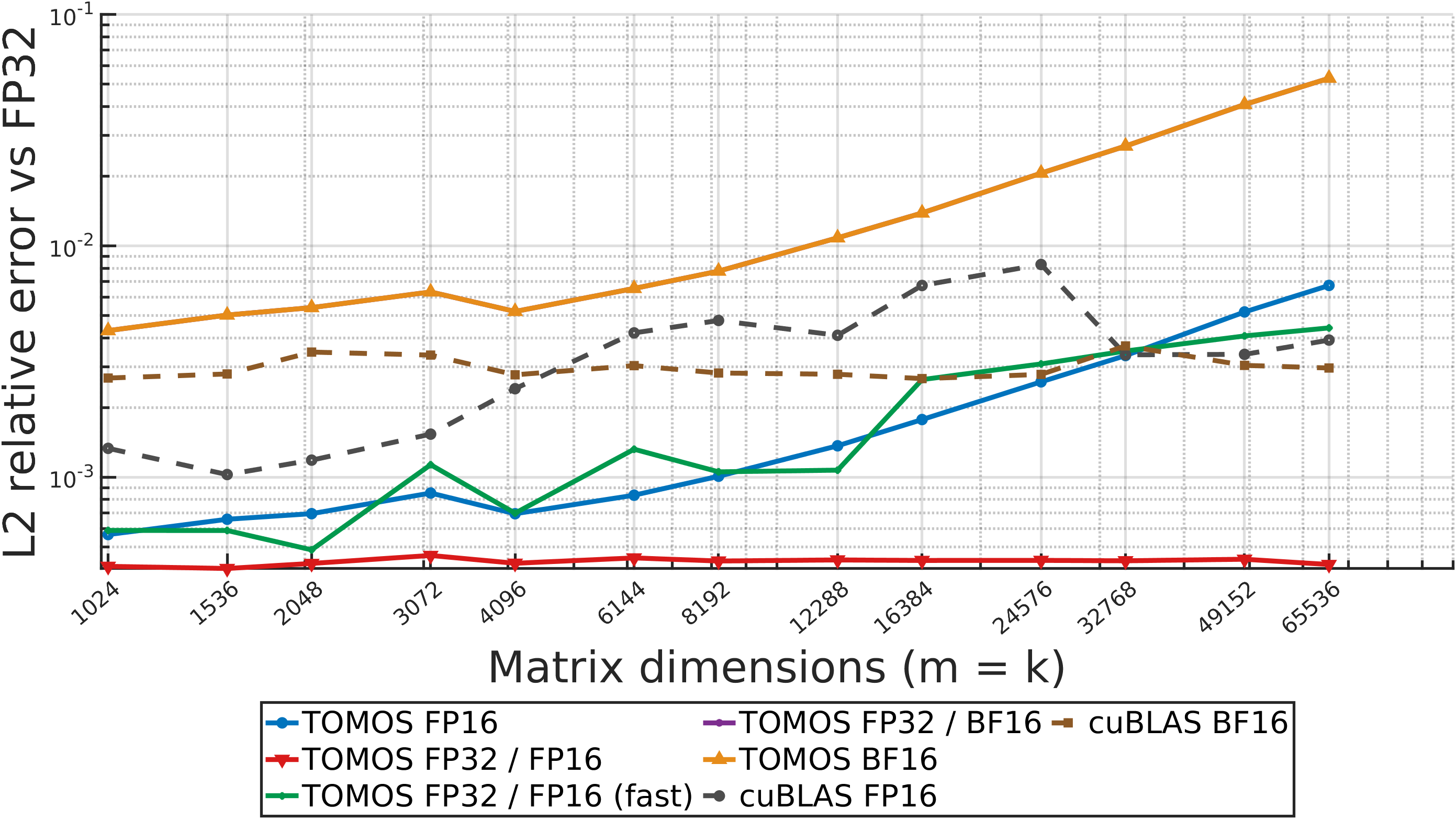}
    \end{subfigure}
        \caption{Relative Frobenius error for 16-bit kernels. Error for \texttt{GEMM} kernels are plotted on the left and \texttt{GEMV} on the right.} 
        \label{error-blas-16}
\end{figure*}

Figures \ref{error-blas-32} and \ref{error-blas-16} depict relative Frobenius error for 32-bit and 16-bit compute kernels respectively. Differences between implementations are expected due to differences in instruction and thread scheduling; despite this all GEMM implementations exhibit the same or improved numerical accuracy to vendor implementations. Error values for GEMV operations contain higher variability. The relative error for ``TOMOS FP16'' GEMV grows asymptotically, but is still consistently smaller than that of cuBLAS and CUTLASS for all but the last data point. The FP32 / FP16 16-bit compute GEMV is considerably more accurate than the other implementations due to the accumulation in FP64. The BF16 GEMV is the only implementation that is less numerically stable.

\subsubsection{Error analysis for magnitude-scaling inputs}
\begin{figure*}[!htbp]
    \centering
    \begin{subfigure}[t]{0.49\textwidth}
        \includegraphics[width=0.85\linewidth]{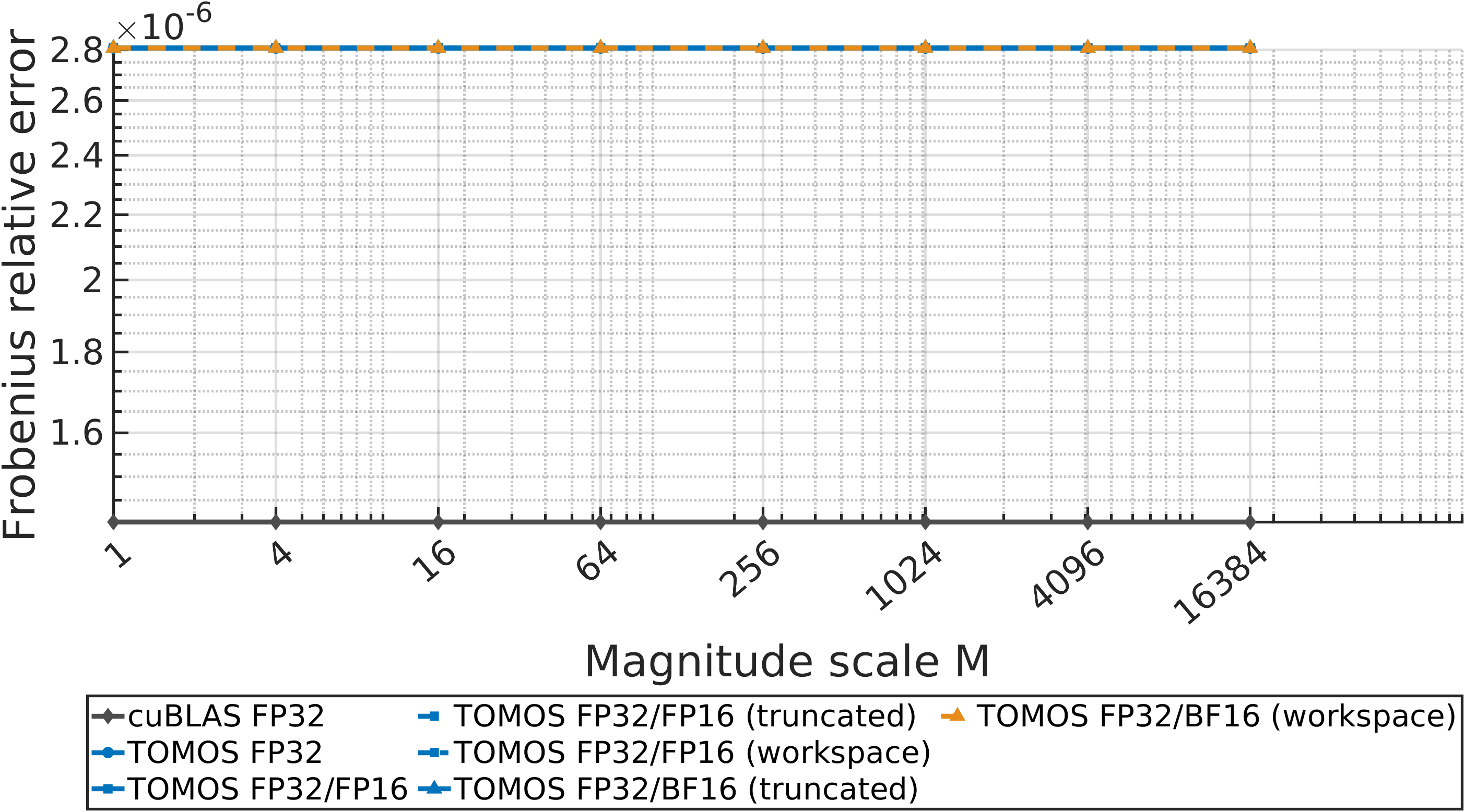}
    \end{subfigure}
    \begin{subfigure}[t]{0.49\textwidth}
        \includegraphics[width=0.85\linewidth]{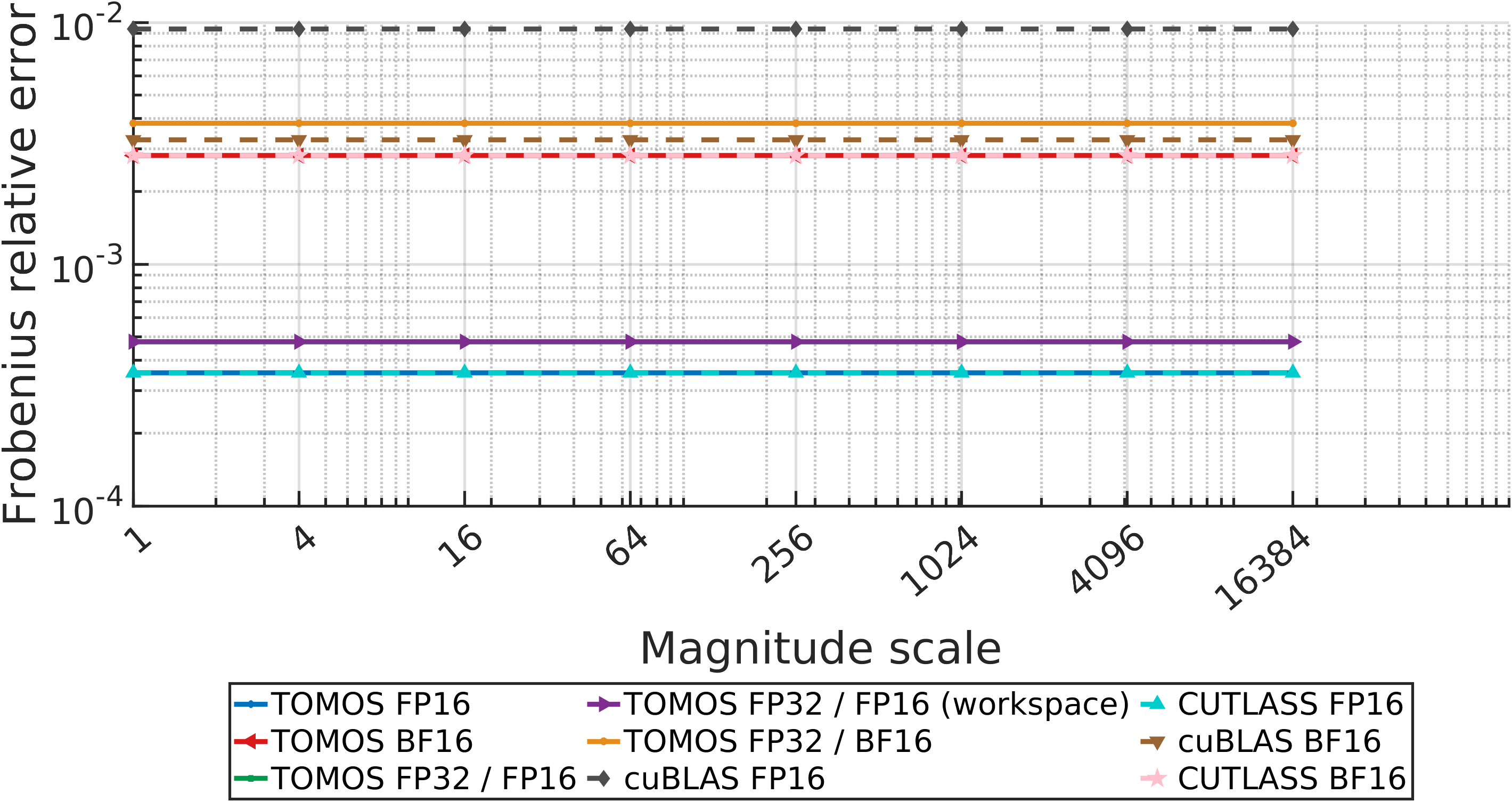}
    \end{subfigure}
        \caption{Relative Frobenius error for GEMM kernels in different compute-precisions, FP32 on the left and FP16 on the right.}
        \label{fig:magnitude-scale}
\end{figure*}

Figure \ref{fig:magnitude-scale} depicts
 error plots for magnitude-scaling experiments on square matrices. Matrix $A$ is sampled from $\text{Uniform}(-M, M)$ and $B$ from $\text{Uniform}(-1\frac{1}{\sqrt{n}}, \frac{1}{\sqrt{n}})$. Results are consistent with those of the dimension-scaling experiment and no variability is exhibited  across different scaling factors.

\subsubsection{Can rounding compensate for the lost mantissa bit}
The FP32/FP16 representation reserves one bit of the primary word for control information, leaving nine mantissa bits rather than the ten available in FP16. We therefore ask whether incorporating information from the discarded bits during encoding can compensate for this loss.

Figure  \ref{fig:scale_1} demonstrates the effect of rounding in 16-bit computations with low-precision TOMOS encodings, showing that
FP16 operations with primary words of the TOMOS FP32 / FP16 representation, labeled as ``Mantissa bits: 9, Sticky bits: 13'',  achieve comparable accuracy with 16-bit encodings initialized under different modes of rounding. The ``Mantissa bits: 10, Sticky bits: 0'' corresponds to 16-bit floating-point entries converted from FP32 via truncation; ``Mantissa bits: 10, Sticky bits: 12'' corresponds to the rounding-to-nearest method applied in FP32-to-FP16 conversions, while the remaining variants  are $9$-mantissa-bit rounding-scheme  approximations. This experiment demonstrates that over  randomized inputs, the primary 9-mantissa bit encodings used by TOMOS, incur smaller numerical penalties than 16-bit encodings constructed by directly truncating 32-bit input words.

\begin{figure}[!htbp]
    \includegraphics[width=0.95\linewidth]{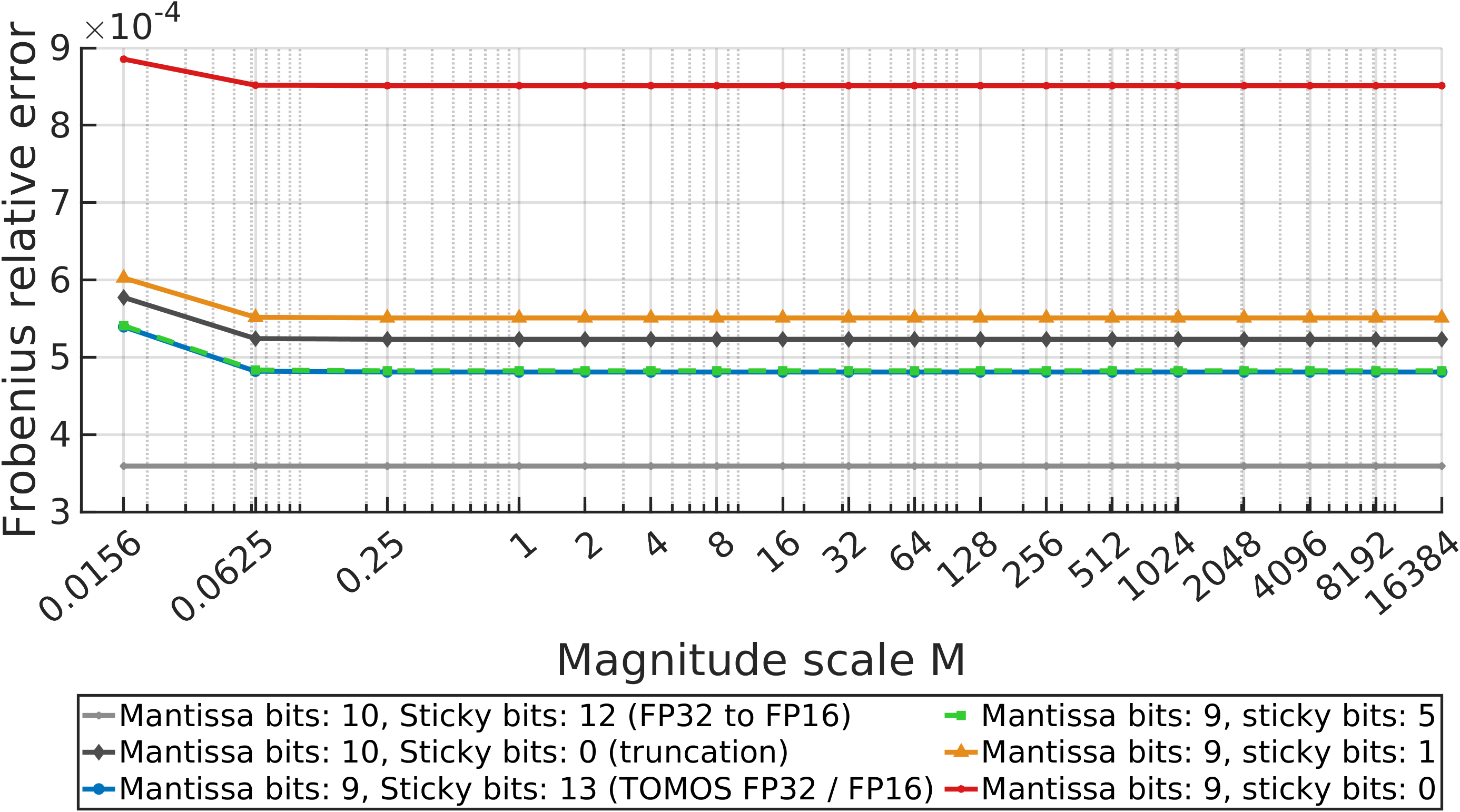}
    \caption{Relative Frobenius error of 16-bit GEMM on square matrices of $49152$ rows and columns initialized in FP32 and converted to FP16.
    (10, 12) rounding to nearest; (10, 0) truncation with 10 mantissa bits; (9, 13) rounding scheme in TOMOS; (9, 0) truncation with $9$ mantissa bits. The (9, 5) and (9, 1) are approximations using fewer sticky bits.}
    \label{fig:scale_1}
\end{figure}

\subsection{End-to-end benefits: Mixed-Precision LSQR}
\label{mixed-precision-lsqr-benchmarks}

Finally, we ask whether the advantages observed at the kernel level translate into an end-to-end numerical solver. In particular, we consider two questions: whether operating on the implicit FP16 view preserves convergence behavior, and whether eliminating the explicit low-precision matrix copy produces a measurable performance and memory benefit.

\begin{figure*}[b]
    \centering
    \includegraphics[scale=0.2]{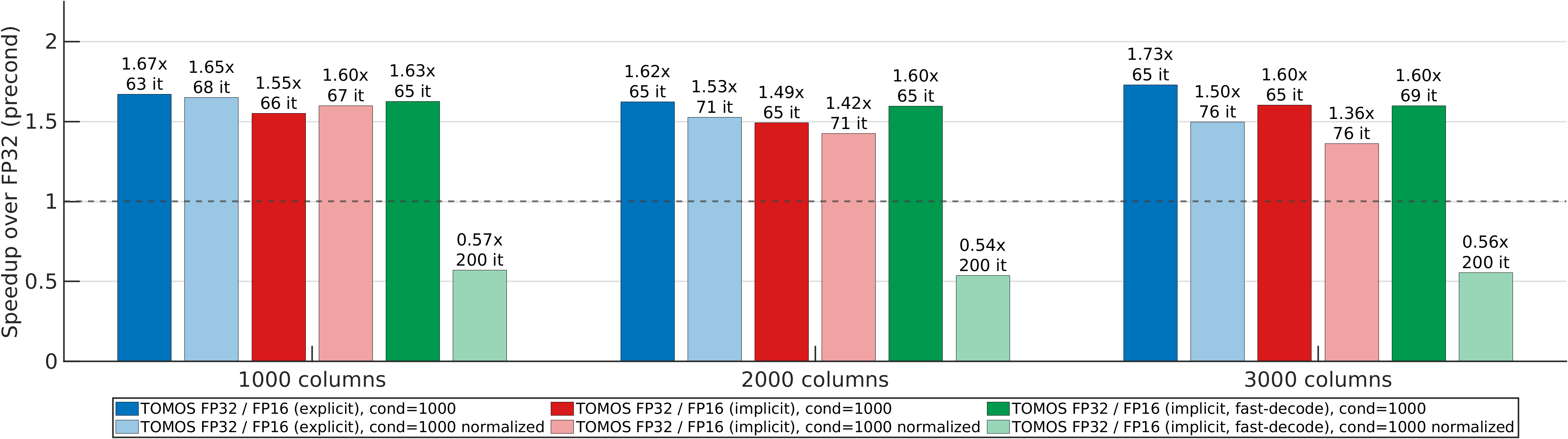}
    \setcounter{figure}{10}
\caption{Speedup plots for mixed precision LSQR implemented using the FP32/16 composite-precision format over FP32.
}
\label{fig:lsqr-perfromance}
\end{figure*}
\begin{figure}[!htpb]
    \includegraphics[width=0.95\linewidth]{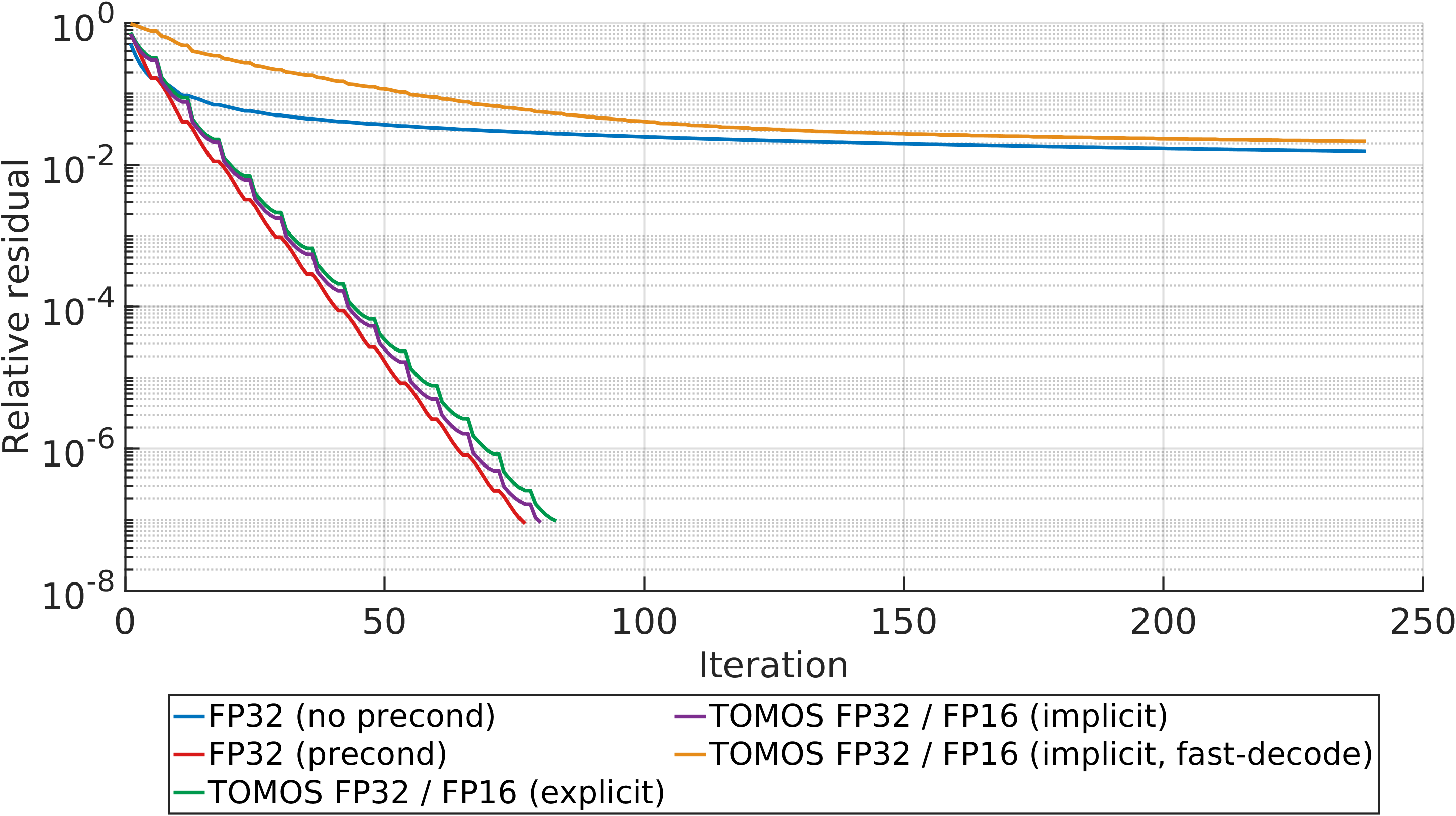} 
    \setcounter{figure}{9}
    \caption{LSQR convergence history of \texttt{matrix-1}, Table \ref{tab:matrices}.
    }
    \setcounter{figure}{11}
    \label{fig:lsqr-convergence}
\end{figure}

Figure \ref{fig:lsqr-convergence} depicts the convergence history of LSQR for \texttt{matrix-1} of Table~\ref{tab:matrices} with $\sim$$18\%$ of its entries stored as subnormals in FP16. The unpreconditioned LSQR iterations in FP32 do not converge, while the preconditioned iterations do. The explicit mixed-precision scheme converges in $70$ iterations while GEMV over FP32 / FP16 storage take an additional $3$ iterations. This is attributed to the fact that primary-word encodings of FP32 / FP16 unavoidably lose one mantissa bit.

On the other hand ``TOMOS FP16 (implicit, fast-decode)''  also operates on the composite format, but stagnates. This happens because fast-decode maps subnormals to zero to improve GEMV throughput, causing LSQR to diverge due to information loss.

Figure~\ref{fig:lsqr-perfromance} reports end-to-end speedup relative to the
FP32 implementation for the two matrices of Table~\ref{tab:matrices}. Bold colors mark results for \texttt{matrix-0} and faded colors for \texttt{matrix-1}. The methods based on explicit copies use blue color; methods using branchless decode over the FP32 / FP16 layout are marked in red and implementations that use fast-decode, to map FP16 underflows to $0$ are shown in green. The number of iterations is similar across explicit (in blue) and implicit, branchless method (in red). Methods relying on fast-decode implementations stagnate for \texttt{matrix-1} due to significant information loss. Converging implementations achieve speedups ranging from $1.42 \times$ to $1.63\times$ over the baseline FP32 implementation. The memory footprint of an LSQR implementation that relies on the FP32 / FP16 memory layout vs an implementation that explicitly stores multiprecision copies is reduced by $\sim33\%$. Importantly, if we consider the low-precision copies as a workspace cost that is necessary to reduce total runtime, \textbf{TOMOS-32/16 kernels achieve workspace reduction of approximately 95\%.}
 
\subsection{Discussion}

Taken together, our experiments establish three properties of the
composite representation. First, its decode overhead can be largely
hidden in compute-bound kernels and substantially reduced in
memory-bound kernels through appropriate decode strategies. Second,
the low-precision view retains numerical accuracy comparable to native
low-precision representations, with rounding compensating for the loss
of one mantissa bit. Third, these kernel-level properties translate
into end-to-end speedups and memory savings in mixed-precision LSQR
without requiring redundant matrix storage.

The results also highlight an important tradeoff in composite-precision
computation. Decode strategies that maximize throughput may sacrifice
numerical information, particularly for subnormal values, whereas
branchless decoding provides more robust behavior at additional
computational cost. Matrix statistics collected during initialization
can therefore inform kernel selection according to the numerical and
performance requirements of the application.

\subsubsection{Comparisons to related work.} TOMOS uses a FP32-sized matrix layout that enables performance advantages within a mixed-precision framework, by exposing computations with a $16$-bit view of a matrix. At the same time, its $32$-bit version may be accessed exactly, for refinement. We do not attempt to improve FP32-performance through algebraic manipulation of the inputs, as is the case for emulation-based methods \cite{mukunoki_2021_accurate_matrix_multiplication, ma_ape_2022, lin_mixpert_2024}. For this reason, the proposed techniques do not exhibit precision loss due to shared-exponent alignment.

\subsubsection{Limitations}
The proposed representations introduce complexity and as such extending the format to a wider range of precision-combinations and GPU architectures, requires careful consideration of hardware-supported formats and  instruction sets. External workspace used to decode tiles in composite storage, also enables portability, since  matrix multiplication routines can operate on the reconstructed batches independently. Our study is inconclusive in regards to portability of GEMV. We also acknowledge that the performance benchmarks of this work were conducted on synthetic data to better control data statistics.

\section{Future Work}
Future work will investigate the applicability of multiprecision encodings in broader domains, including deep neural network and scientific-computing workloads. We will explore automated optimization to support additional precision-combinations and adaptive, variable-precision computing. Lastly, we will explore software / hardware co-design paradigms for composite-precision computations.

\bibliographystyle{IEEEtran}
\bibliography{IEEEabrv,bibliography.bib}

\end{document}